\documentclass[%
longbibliography,
reprint,
superscriptaddress,
amsmath,amssymb,
aps,
prx,
floatfix
]{revtex4-2}

\usepackage{graphicx}
\usepackage{dcolumn}
\usepackage{bm}
\usepackage{hyperref}
\hypersetup{
    colorlinks=true,
    linkcolor=blue,
    filecolor=magenta,      
    urlcolor=cyan,
    pdftitle={Overleaf Example},
    pdfpagemode=FullScreen,
    }

\usepackage{subfloat}
\usepackage{graphicx}
\usepackage{subcaption}
\usepackage{braket}
\usepackage{xcolor}

\begin{document}


\title{Label-free quantum super-resolution imaging using entangled multi-mode squeezed light}

\author{Daniel Soh}
\email{dbsoh@sandia.gov}
\affiliation{Sandia National Laboratories, Livermore, California 94550
}
\author{Eric Chatterjee}
\email{ericchatt5@gmail.com}
\affiliation{Sandia National Laboratories, Livermore, California 94550
}

\date{\today}

\begin{abstract}

In this study, we explore the theoretical application of entangled multi-mode squeezed light for label-free optical super-resolution imaging. By generating massively entangled multi-mode squeezed light through an array of balanced beam splitters, using a single-mode squeezed light input, we create a multi-mode quantum light state with exceptional entanglement and noise suppression below the shot noise level. This significantly reduces imaging measurement errors compared to classical coherent state light imaging when the same number of photons are used on the imaging sample. We demonstrate how to optimize the imaging system's parameters to achieve the Heisenberg imaging error limit, taking into account the number of entangled modes and photons used. We also examine the effects of optical losses in the imaging system, necessitating adjustments to the optimized parameters based on the degree of optical loss. In practical applications, this new quantum imaging approach reduces the number of photons needed to achieve the same image quality by two orders of magnitude compared to classical imaging methods that use non-entangled, non-squeezed coherent state light.

\end{abstract}

\maketitle


\section{Introduction}

Super-resolution imaging is a technique showing an object's visual features smaller than the so-called Rayleigh diffraction limit, which is approximately on the order of probing light's wavelength. Typical \textit{classical} (as opposed to \textit{quantum}) super-resolution imaging relies on a labeling technique where optically sensitive materials (dyes) are injected into samples, and either active or passive fluorescence imaging techniques are used to localize the optically responding labels, revealing nanoscale features. This kind of technique includes photoactivated localization microscopy (PALM), stochastic optical reconstruction microscopy (STORM), stimulated emission depletion (STED) microscopy, and structured illumination microscopy (SIM) (for a comprehensive review on these techniques, see \cite{cremer2017super, poole2021optical}). However, these \textit{classical} techniques have shortfalls, including limited target biomarkers (only a few at a time) and enormously long imaging acquisition time. Most of all, sample preparation (labeling) is time-consuming, and the toxicity of dyes on the benign imaging samples is not well understood. 

There are other non-optical imaging methods that allow even a sub-nanometer resolution. These include electron microscopy (EM) \cite{de2011electron} and soft x-ray tomography (SXT) \cite{parkinson2008quantitative}. However, these use high energy particles as imaging probes, which may result in undesired side-effects: the high-energy particles may disturb the sample's quantum states (e.g., ionization, elevation to higher energy states) or even destroy fragile samples such as sensitive chemical molecules or biosamples. For these sensitive samples, optical imaging provides an excellent alternative method due to the low energy optical photons leading to less-interfering imaging, which however is limited in imaging resolution due to the finite size of photon wavepacket. 

Many attempts were made for label-free optical imaging. Recognizing that the ultimate imaging resolution is limited by the quantum fluctuation of light (i.e., shot noise) \cite{kolobov2000quantum}, multiple research results focused on how to reduce the effect of quantum noise in the probing light \cite{beskrovnyy2005quantum, wang2012quantum}. These included the usage of quantum entanglement of single photons \cite{giovannetti2004quantum, sajia2021super} and squeezed light \cite{treps2005quantum, embrey2015observation, treps2004nano, feng2013long}. More recently, quantum sensing in general observed a broader use of the combination of quantum entanglement and squeezing in order to enhance the signal-to-noise ratio of measurement under the condition of low photon numbers \cite{zhuang2018distributed, zhang2021distributed}. In this context, a single-mode squeezed light is used to generate massively entangled multi-mode squeezed light (MEMSL) that bears the excellent feature of reducing measurement errors in the Heisenberg limit with respect to the number of entangled modes \cite{guo2020distributed}. So far, the related research has focused on measurements of lumped elements. Inspired by this new development of the combination of quantum entanglement and squeezing, we present for the first time a new quantum imaging technique that is label-free utililzing the MEMSL as probe light to enhance the image resolution. MEMSL imaging has a superior performance with a significantly reduced imaging error in the Heisenberg limit where the imaging error scales as $1/MN$ ($M$: number of entangled modes in MEMSL, $N$: number of photons in each mode), rather than the standard limit $1/\sqrt{MN}$ of classical imaging utilizing non-entangled coherent light imaging repeating $M$-times. While one can achieve super-resolution imaging even with a classical method using a coherent state of light, provided that sufficient averaging reduces the measurement noise, the new MEMSL imaging accomplishes the same level of imaging resolution with much reduced requirement for averaging, which therefore accomplishes significantly shorter imaging acquisition time with a much smaller number of photons used. 

There are many cases where the number of photons impinging on the imaging sample must be limited: quantum state tomography of molecules requires only a small number of probing photons in order not to disturb the quantum state to be measured \cite{zhang2021quantum, mouritzen2006quantum, skovsen2003quantum}. Bioimaging, in general, requires only a small number of photons in order not to disturb the bioprocesses or overheat the fragile biomolecules \cite{voronin2010ionization}. Furthermore, chemical surface reactive intermediate molecules are so fragile that excessive probe light may dissociate or ionize them, making them hard to detect \cite{cremer1996ethylene}. All these cases demand limiting the number of photons impinging on the sensitive and fragile imaging samples. Hence, it is imperative to study how one can accomplish the best imaging resolution under such a strict limitation on number of photons impinging on the samples. Our new MEMSL imaging addresses this important question by providing a breakthrough quantum imaging utilizing entanglement and squeezing.

In this paper, we start with introducing MEMSL and how it is generated, which is followed by the theory of quantum MEMSL imaging that is compared to two other cases, namely, non-entangled single-mode squeezed light imaging repeating multiple times and \textit{classical} imaging using coherent state light repeating multiple times. Then, we discuss the obtainable image resolution with a limited number of photons. We also discuss the impact of optical losses in the imaging system.  We then provide numerical case studies and discuss how effective MEMSL imaging is in realistic situations.

\section{Quantum Imaging Theory}

\begin{figure}[!tb]
	\includegraphics[width=0.35\textwidth]{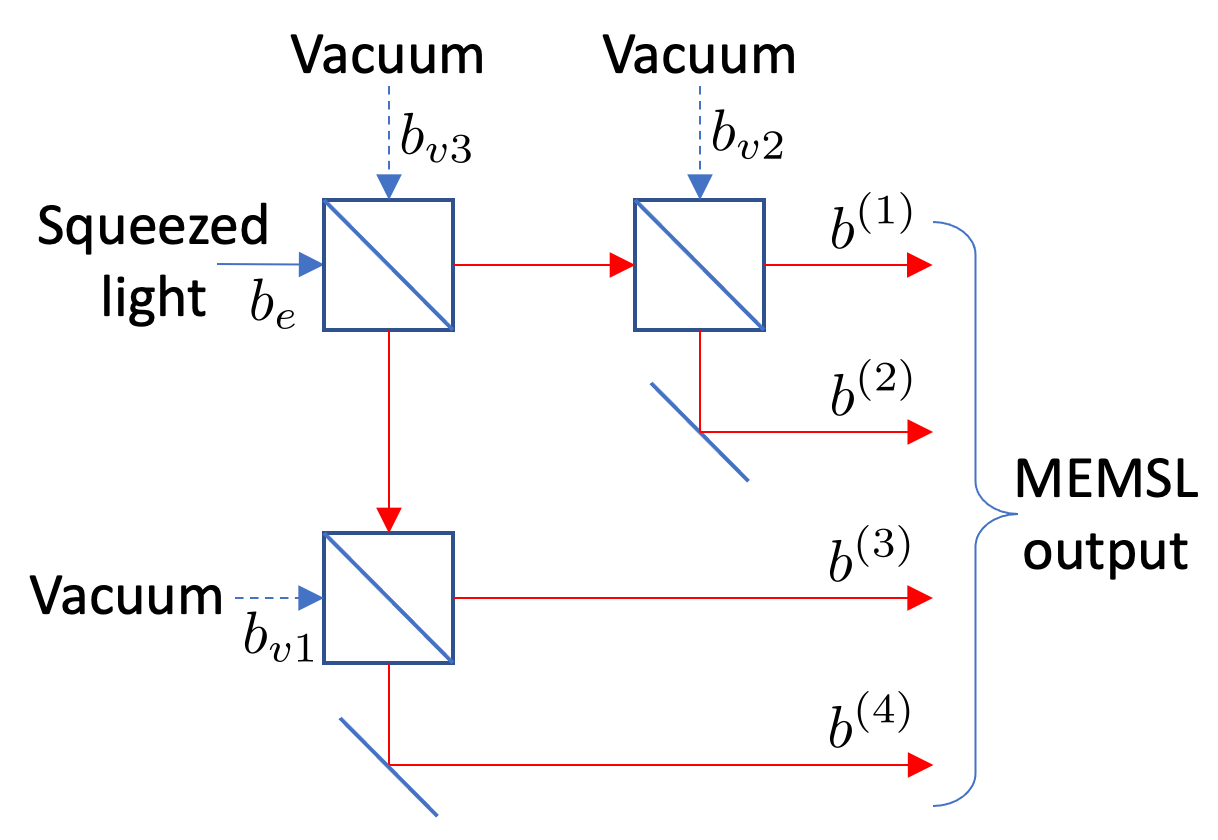}
	\caption{A method to generate a 4-mode MEMSL using an array of balanced beam splitters with an input from a single-mode squeezed light.} \label{fig:MEMSL}
\end{figure}

In this section, we briefly introduce MEMSL that will play an important role in our scheme and discuss how one can utilize MEMSL as a probe for quantum imaging. In order to create the MEMSL light state, one starts with a single-mode squeezed light. A network of beam-splitters produces a multi-mode output when only one input is seeded by the single-mode squeezed light while all the other inputs are in vacuum state \cite{zhuang2018distributed, zhang2021distributed}. For example, let us consider a 4-mode MEMSL as shown in Fig. \ref{fig:MEMSL}, which uses three 50-50 beam splitters. It is well known that the multi-mode output is quantum mechanically entangled, resulting in a relation \cite{guo2020distributed}
\begin{equation}
	b_e = \frac{1}{2} (b^{(1)} + b^{(2)} + b^{(3)} + b^{(4)}),
\end{equation}
where $b^{(i)}$ represents the annihilation operator of the $i$th output port mode and $b_e$ is the annihilation operator of the input single-mode squeezed light. The above equation has a semi-classical interpretation that the field injected in the squeezed light input port is divided into the four output modes equally. Quantum mechanically, the photons in the input squeezed mode are divided into four modes with equal probabilities. In general, it can be shown that the following holds for MEMSL with $M$ modes \cite{guo2020distributed}:
\begin{equation}
	b_e = \frac{1}{\sqrt{M}} \sum_{m=1}^M  b^{(m)}. \label{eq:MEMSL-eq}
\end{equation}

\begin{figure}[!tb]
	\includegraphics[width=0.3\textwidth]{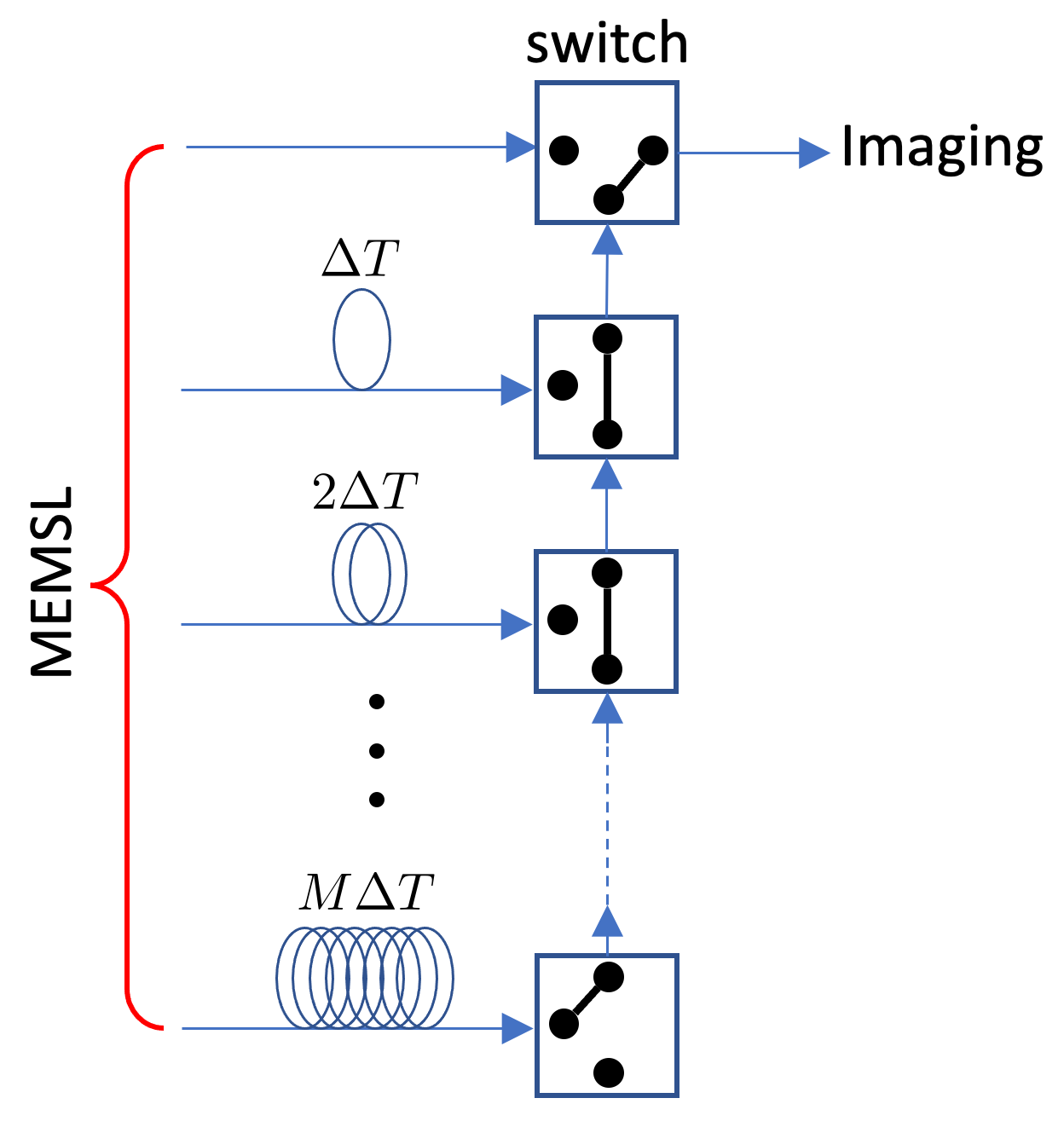}
	\caption{A time-delayed feeding scheme for MEMSL's individual modes to the imaging station.} \label{fig:time-delayed-feed}
\end{figure}

The multiple modes in MEMSL with M modes are used in the imaging scheme one mode at a time (Fig. \ref{fig:time-delayed-feed}). In this scheme, each output mode of $M$-mode MEMSL is selected sequentially using a switching array with differential time delays. Hence, the $m$th mode will be fed into the imaging station after $m \Delta T$ time delay where $\Delta T$ is appropriately set between two adjacent measurements in the imaging station. The measurement is taken for each of the $m$ modes, and later processed together.

\begin{figure}[!tb]
	\includegraphics[width=0.4\textwidth]{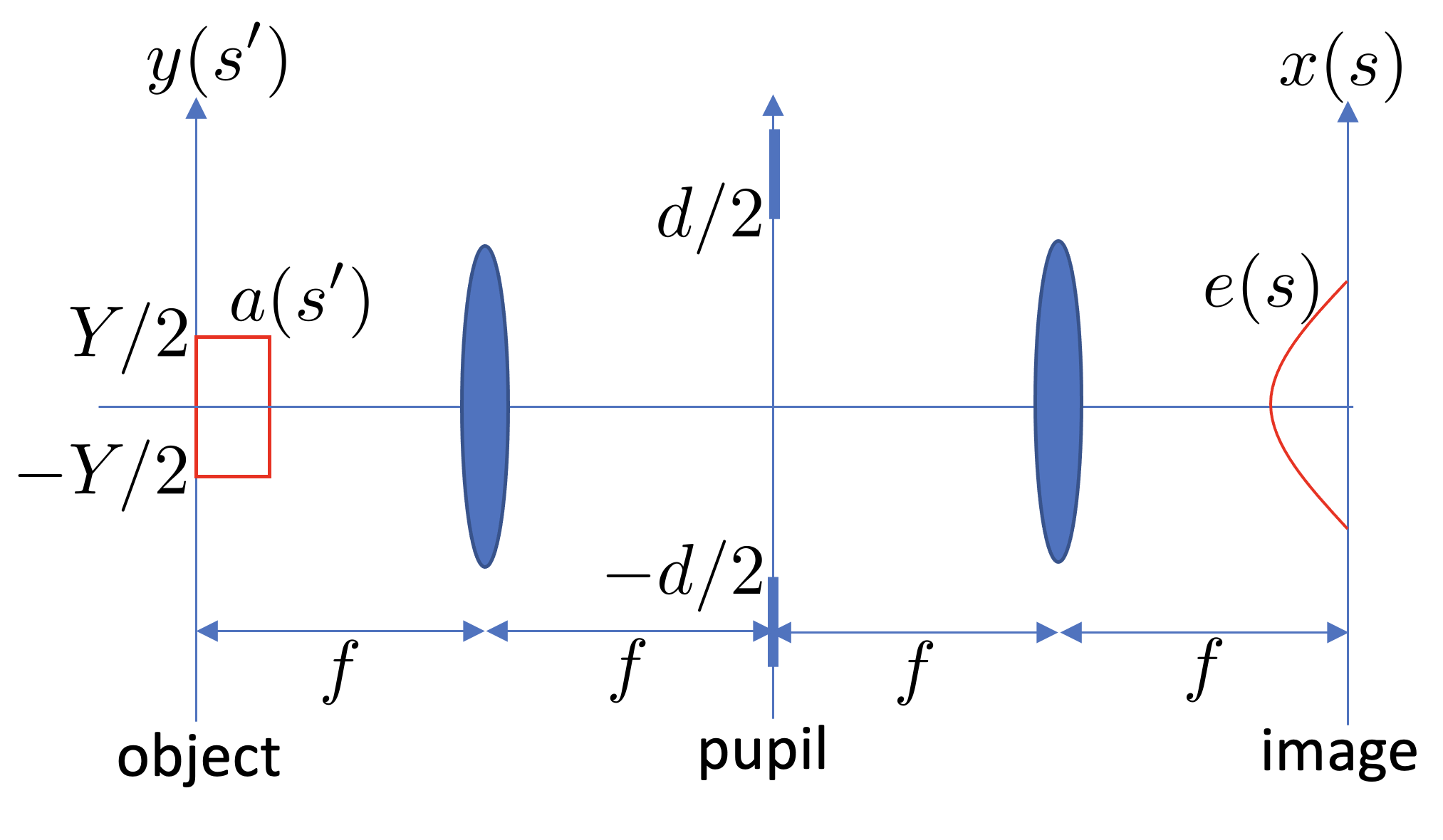}
	\caption{A one-dimensional 4-$f$ imaging system.} \label{fig:imaging}
\end{figure}

Following the treatment in Kolobov and Fabre \cite{kolobov2000quantum}, we will consider a one-dimensional imaging system shown in Fig. \ref{fig:imaging} where an object with size $Y$ is placed in the object plane (coordinate $y$). Generalization of MEMSL imaging to higher dimensions is rather straightforward (see Section  \ref{sec:disc}. Discussions and conclusions for more details). A pair of lenses with focal length $f$ and pupil size $d$ are used to image the object on the image plane (coordinate $x$). Since the lenses have only a finite pupil size, higher spatial-frequency components in the pupil plane are lost. As a result, the image on the image plane has smooth boundaries comprising of only lower spatial-frequency components \cite{kolobov2000quantum}. Let us normalize the coordinates such that $s = 2 x / Y, s' = 2 y /Y$. This normalization allows the generalization of the theory to other two-lens imaging systems with different focal lengths where $s' = 2 y /(Yf'/f)$ with $f'$ representing the focal length of the second (imaging plane side) lens. We will also define a space-bandwidth product
\begin{equation}
	c = \frac{\pi d Y}{2 \lambda f}, \label{eq:spatial-bandwidth}
\end{equation}
where $\lambda$ is the wavelength of the light. 

In the following, we will introduce three imaging methods, namely, (1) MEMSL imaging, (2) single-mode squeezed light imaging repeating $M$ times, and (3) \textit{classical} imaging using coherent-state light repeating $M$ times, in order to demonstrate the superior performance of MEMSL imaging over the two others.

\subsection{MEMSL imaging method}

We emphasize that we extend the quantum imaging much further by introducing the optimization of squeezing and displacement parameters in the context of applying the MEMSL technique, while the Kolobov and Fabre's results described merely the advantage of using squeezed light in imaging \cite{kolobov2000quantum}. Our imaging method accomplishes a Heisenberg imaging error reduction with respect to the number of photons impinging on the imaging sample and the number of modes in MEMSL, which is a significant milestone for general quantum imaging. 

Let us assume a plane wave for each of the MEMSL output modes. Our imaging scheme measures any combination of light phase and amplitude using a homodyne or a heterodyne measurement (or any superposition between the two) after the probing light interacts with the object. We consider, without losing generality of analysis, the case where the object's spatial information is imprinted on the optical phase of the interacting light. The case where the object's spatial information is imprinted in the amplitude of light (or any combination of amplitude and phase) can be treated similarly, just by rotating the squeezing angle of the squeezed light in the phase plane appropriately for minimizing the quantum noise in the desired light quadrature angle.

We apply a MEMSL treatment and finally reconstruct the objective in a new way that fully utilizes the quantum entanglement in MEMSL to achieve the minimal error in the image reconstruction (see the general imaging treatment and the definitions of prolate spheroidal functions $\varphi_j(s')$, $\chi_j(s')$, and $\psi(s)$ in Appendix \ref{sec:general-imaging}). For this, let's define the estimate object field (in an orthogonal decomposition form) $\tilde{a}^{\mathrm{avg}} (s') = \sum_{j=0}^Q \tilde{a}_j^\mathrm{avg} \varphi_j (s')$ for $|s'|\leq 1$. Here, we limit the summation to $j=Q$, which is the highest order of useful prolate spheroidal wave functions. The orders higher than $Q$ do more harm than good because the quantities $\langle \tilde{a}_{j>Q}^\mathrm{avg} \rangle$ are noisy. The maximum order number $Q$ is determined by both quantum and measurement noise. It will be shown later that $Q$ is directly related to the obtainable resolution of the reconstructed object field because higher orders of the prolate spheroidal wave functions provide faster spatial variation (oscillation) of the object field features (i.e., a better imaging resolution of the reconstructed object field is obtained if higher orders of prolate spheroidal functions are added).

We now show that an average of $e_j^{(m)}$ over all values of the mode index $m$ in MEMSL will accomplish the minimum imaging error. Eq. \eqref{eq:tilde-ae} results in a simple relation for the estimate $\tilde{a}_j^\mathrm{avg} = (1/\sqrt{\lambda_j}) e_j^\mathrm{avg}$ where $e_j^\mathrm{avg} = (1/M) \sum_{m=1}^M e_j^{(m)}$: the estimate object field for $|s'|\le1$ is
\begin{align}
	\tilde{a}^\mathrm{avg} (s') = \sum_{j=0}^Q \tilde{a}_j^\mathrm{avg} \varphi_j(s') = \sum_{j=1}^Q \frac{e_j^\mathrm{avg} }{\lambda_j} \psi_j (s'). \label{eq:a-const}
\end{align}

Again, the ultimate goal of imaging is to obtain $\phi(s')$, which is the spatially-dependent phase distribution (Eq. \eqref{eq:a-b}). We note the following crucial relation between $\tilde{a}^\mathrm{avg} (s')$ and $\phi (s')$, which is derived using the characteristic MEMSL relation in Eq. \eqref{eq:MEMSL-eq}:
\begin{equation}
	\tilde{a}^\mathrm{avg} (s') = \frac{1}{M} \sum_{m=1}^M b^{(m)} e^{i \phi (s')} = \frac{1}{\sqrt{M}} b_e e^{i \phi (s')}. \label{eq:a-MEMSL}
\end{equation}
Combed with Eq. \eqref{eq:a-const},  this relation extracts $\phi (s')$ from the measured $e_j^\mathrm{avg}$. 

Let us now assume that $\phi(s')$ is sufficiently small for all $|s'|\leq1$. We introduce the quadrature operators for $b_e$ as 
\begin{align}
	b_e &= b_{e1} + i b_{e2}, \nonumber \\
	b_{e1} &= (b_e + b_e^\dagger)/2, \nonumber \\
	b_{e2} &= (b_e - b_e^\dagger)/(2i).
\end{align}
The quadrature operators are Hermitian. We also introduce the quadrature operators for $b^{(m)}$ similarly: $b_1^{(m)} = (b^{(m)} + b^{(m)\dagger})/2$ and $b_2^{(m)} = (b^{(m)} - b^{(m)\dagger})/(2i)$. We assume a displaced squeezed state for $b_e$-field such that
\begin{align}
	&\langle b_{e1} \rangle = \sqrt{2} \alpha_e, \quad \langle b_{e2} \rangle = 0, \nonumber \\
	&\langle (\Delta b_{e1})^2 \rangle = \frac{e^{+2 r_e}}{4}, \quad \langle (\Delta b_{e2})^2 \rangle = \frac{e^{-2 r_e}}{4}, \label{eq:be12_stat}
\end{align}
where $\alpha_e$ is the displacement and $r_e$ is the squeezing parameter. We note that this state is a ``phase squeezed'' state, suitable for phase estimate. Then, one obtains from Eq. \eqref{eq:a-MEMSL}
\begin{equation}
	\langle \tilde{a}_2^\mathrm{avg} (s') \rangle = \frac{\sqrt{2} \alpha_e }{\sqrt{M}} \sin \phi (s')  \approx \alpha_e  \phi (s') \sqrt{\frac{2}{M}}, \label{eq:a_avg_phi}
\end{equation}
where $\tilde{a}_2^\mathrm{avg} (s') = (\tilde{a}^\mathrm{avg} (s') - \tilde{a}^\mathrm{avg \dagger} (s'))/(2i)$ is a quadrature operator. Based on these and Eq. \eqref{eq:a-const}, we define the estimate $\tilde{\phi} (s')$, which is a Hermitian quantum field operator, as following:
\begin{align}
	\tilde{\phi} (s') &= \frac{1}{M} \sum_{m=1}^M \sum_{j=0}^Q \frac{1}{\alpha_e \lambda_j} \sqrt{\frac{M}{2}} \psi_j (s') e^{(m)}_{j2} \nonumber \\
	&= \frac{1}{\alpha_e \sqrt{2M}} \sum_{j=0}^Q \frac{\left( \sum_{m=1}^M e^{(m)}_{j2} \right)}{\lambda_j} \psi_j (s'), \label{eq:recon-recipe}
\end{align}
where $e_{j2}^{(m)} = (e^{(m)}_j - e^{(m)\dagger}_j)/(2i)$, representing the imaginary part of the image component created by the light in mode $m$ and corresponding to the $j$th spheroidal function. 

The goal of imaging is to obtain the best estimate $\langle \tilde{\phi}(s') \rangle$ under the influence of the quantum noise embedded in $e_{j2} ^{(m)}$. This noise in turn relates to the noise embedded in the source photon field $b_2^{(m)}$ through the following (see Appendix~\ref{sec: Derivation of Quantum Noise} for the derivation):
\begin{equation} \label{eq:ej2}
	e^{(m)}_{j2} = b_2^{(m)} \int_{-\infty}^\infty ds' \psi_j(s) + b_1^{(m)} \int_{-1}^1 ds' \phi (s') \psi_j (s'),
\end{equation}
leading to the following for the composite imaginary image component for the $j$th spheroidal function:
\begin{align}
	&\sum_{m=1}^M e^{(m)}_{j2} = \nonumber \\
	&  b_{e2} \sqrt{M} \int_{-\infty}^\infty ds' \psi_j (s') + b_{e1} \sqrt{M} \int_{-1}^1 ds' \phi  (s') \psi_j (s'). \label{eq:eb}
\end{align}
The noise in the reconstructed object field, resulting from the noise in the image field, becomes the following (see Appendix~\ref{sec: Derivation of Quantum Noise}):
\begin{align}
\begin{split}
	\sigma &= \sqrt{\int_{-\infty}^{\infty} ds' ~ \langle (\Delta \tilde{\phi} (s'))^2 \rangle} \\
    &\approx \frac{e^{-r_e}}{2\sqrt{2} \alpha_e} \sqrt{\sum_{j=0}^Q \frac{A_j^2}{\lambda_j^2}} ,
\end{split}
\end{align}
which is the root-mean-squared sum of all noises in the reconstructued spatially-dependent phase. The coefficients $A_j$ are defined as follows:
\begin{equation}
	A_j = \left|\int_{-\infty}^\infty ds' \psi_j (s') \right|, \label{eq:Aj}
\end{equation}
Note that the reconstructed-object-field noise associated with a given basis function $j$ scales inversely with $\lambda_j$. This is because for a lower $\lambda_j$, the object-to-image attenuation is more pronounced. As such, when deriving the reconstructed object field from the image field, a greater amplification is required, which also causes significant amplification of the noise from the image field. Based on the fact that $\lambda_j$ drops rapidly with increasing $j$, we can make the approximation $A_Q^2 / \lambda_Q^2 \gg \sum_{j=1}^{Q-1} A_j^2 / \lambda_j^2$, which is numerically verifiable. This leads to
\begin{equation}
	\sigma \approx \frac{e^{-r_e} A_Q}{2 \sqrt{2} \alpha_e \lambda_Q}. \label{eq:lossless-error}
\end{equation}

We now consider the fact that the number of photons impinging on the sample must be limited. The average number of photons hitting the sample in each mode of MEMSL is given as
\begin{equation}
	N_e = \frac{1}{M} (\alpha_e^2 + \sinh^2 r_e), \label{eq:photon-number}
\end{equation}
since the photons in the single-mode squeezed light are equally divided into the $M$ modes of MEMSL. 

Now, we introduce the optimization for obtaining the minimum imaging error $\sigma$ by varying the squeezing parameter $r_e$ and the displacement $\alpha_e$ while fixing $N_e = N$, for a given number of modes $M$ in the MEMSL. Substituting Eq.~\eqref{eq:photon-number} into Eq.~\eqref{eq:lossless-error}, we can express the amplitude-squared of the noise, i.e. $\sigma^2$, in terms of the squeezing parameter $r_e$ as follows:
\begin{equation}
\sigma^2 \approx \frac{A_Q^2}{8 \lambda_Q^2} \bigg(\frac{e^{-2r_e}}{MN - \sinh^2{r_e}}\bigg).
\end{equation}
Taking the derivative of $\sigma^2$ with respect to $e^{-2r_e}$ and setting it to zero (corresponding to the minimum value for $\sigma^2$), the optimal solution set is easily obtained to be
\begin{equation}
	e^{-r_e} = \frac{1}{\sqrt{1 + 2 MN}}, \quad \alpha_e = \frac{\sqrt{MN(1+MN)}}{\sqrt{1 + 2 MN}}. \label{eq:optimal_solution_MEMSL}
\end{equation}
The tradeoff can be conceptualized as follows: The phase uncertainty per photon (i.e., the angle subtended by the second-quadrature uncertainty in the phase space distribution) approximately equals the ratio between the second-quadrature uncertainty and the amplitude. As the squeezing parameter $r_e$ is increased, the second-quadrature uncertainty per photon decreases as desired, but the amplitude per photon also decreases. Consequently, if the light is too strongly squeezed, the phase uncertainty per photon (i.e., the phase noise) will start to increase, requiring optimization. With the above solution, the optimal (minimal) imaging error $\sigma$ is
\begin{equation}
	\sigma^{\mathrm{opt}} = \left( \frac{A_Q}{2 \sqrt{2} \lambda_Q} \right) \frac{1}{\sqrt{MN ( 1 + MN)}}. \label{eq:MEMSL-noise}
\end{equation}
We note that, for a large $MN$, the error scales as $\sim 1/MN$, which is a Heisenberg limit with respect to both $M$ and $N$. 

\subsection{Non-entangled squeezed light imaging with $M$-time repeating measurements}

To compare the MEMSL imaging technique above with other alternatives, we introduce non-entangled single-mode squeezed light imaging. In this method, a single squeezed light is used for $M$ times and the result is averaged. 

Let us designate the annihilation operator for each non-entangled single-mode squeezed light as $b_e^{(m)} = b_{e1}^{(m)} + i b_{e2}^{(m)}$ for $m=1, 2, \cdots, M$ with $b_{e1}^{(m)} = (b_e^{(m)} + b_e^{(m)\dagger})/2$ and $b_{e2}^{(m)} = (b_e^{(m)} - b_e^{(m)\dagger})/(2i)$. The statistics of the quadrature operators is
\begin{align}
	&\langle b_{e1}^{(m)} \rangle = \sqrt{2} \alpha'_e, \quad \langle b_{e2}^{(m)} \rangle = 0, \nonumber \\
	&\langle (\Delta b_{e1}^{(m)})^2 \rangle = \frac{e^{+2 r'_e}}{4}, \quad \langle (\Delta b_{e2}^{(m)})^2 \rangle = \frac{e^{-2 r'_e}}{4},
\end{align}
It is straightforward to derive
\begin{equation}
	\tilde{a}_2^\mathrm{avg} (s') \approx  \sqrt{2} \alpha'_e \tilde{\phi} (s') ,
\end{equation}
which leads to
\begin{equation}
	\tilde{\phi} (s') = \frac{1}{\alpha_e M \sqrt{2}} \sum_{j=0}^Q \frac{\left( \sum_{m=1}^M e_{j2}^{(m)}\right)}{\lambda_j} \psi_j (s'),
\end{equation}
\begin{align}
	\sum_{m=1}^M e_{j2}^{(m)} = &\left( \sum_{m=1}^M b_{e2}^{(m)} \right) \int_{-\infty}^\infty ds' \psi_j (s') \nonumber \\
	&+ \left( \sum_{m=1}^M b_{e1}^{(m)} \right) \int_{-1}^1 ds' \tilde{\phi} (s') \psi_j (s'),
\end{align}
and finally, one obtains
\begin{equation}
	\sigma \approx \frac{e^{-r'_e} A_Q}{2 \sqrt{2 M} \alpha'_e \lambda_Q}. \label{eq:sigma-single-squeezed}
\end{equation}
This implies that the error in this case scales as $\sim 1/\sqrt{M}$, which is a well known result for an average over multiple non-entangled measurement trials. Also, we note that the photon number in each trial amounts to $N_e = {\alpha'_e}^2 + \sinh^2 r'_e$. Substituting this into Eq.~\eqref{eq:sigma-single-squeezed}, $\sigma^2$ becomes
\begin{equation}
\sigma^2 \approx \frac{A_Q^2}{8M\lambda_Q^2} \bigg(\frac{e^{-2r'_e}}{N - \sinh^2{r'_e}}\bigg)
\end{equation}
The optimal solutions of $\alpha'_e, r'_e$ that minimize $\sigma^2$ are
\begin{equation}
	e^{-r'_e} = \frac{1}{\sqrt{1 + 2N}}, \quad \alpha'_e = \frac{\sqrt{N(1+N)}}{\sqrt{1 + 2 N}},
\end{equation}
which leads to the optimal (minimal) imaging error
\begin{equation}
	\sigma^\mathrm{opt} = \left( \frac{A_Q}{2 \sqrt{2} \lambda_Q} \right)  \frac{1}{\sqrt{M} \sqrt{N(1 + N)}}. \label{eq:ne-squeezing}
\end{equation}
We note that for large $N$, the error scales as $\sim 1/(N\sqrt{M} )$, which is a standard limit with respect to $M$ but a Heisenberg limit with respect to $N$. This is due to the fact that in the absence of entanglement, squeezing degrades the amplitude per photon more rapidly, with the rate of degradation with respect to squeezing increasing with $M$. Consequently, without entanglement, the noise decreases more slowly with the number of modes $M$. This result reveals that the entanglement in different modes of MEMSL is essential to make the $M$-dependence be a Heisenberg limit. 

\subsection{Non-entangled coherent state light imaging with $M$-time repeating measurements}

Next, we calculate the imaging error when \textit{classical} non-entangled non-squeezed coherent state light is used for $M$ times and the result is averaged. 

This case is quite similar to the previous case of non-entangled squeezed light imaging except $r'_e = 0$. We designate the annihilation operator of the individual coherent-state light as $b_e^{(m)} = b_{e1}^{(m)} + i b_{e2}^{(m)}$ where
\begin{align}
	&\langle b_{e1}^{(m)} \rangle = \sqrt{2} \alpha'_e, \langle b_{e2}^{(m)} \rangle = 0, \nonumber \\
	&\langle (\Delta b_{e1}^{(m)})^2 \rangle = \langle (\Delta b_{e2}^{(m)})^2 \rangle = \frac{1}{4},
\end{align}
it is straightforward to obtain, replacing $r'_e = 0$ in Eq. \eqref{eq:sigma-single-squeezed}:
\begin{equation}
	\sigma \approx \frac{A_Q}{2 \sqrt{2M} \alpha'_e \lambda_Q}.
\end{equation}
Then, with the fact that the number of photons in individual imaging trials is $N_e = {\alpha'_e}^2 = N$, the above turns into
\begin{equation}
	\sigma \approx \left( \frac{A_Q}{2 \sqrt{2} \lambda_Q} \right) \frac{1}{\sqrt{MN}}, \label{eq:sigma-coherent-light}
\end{equation}
which implies that the error scales as the standard limit for both $M$ and $N$. Note that there is no need to optimize $\alpha'_e$ here because it is set by the number of photons used in each measurement. It is also worth noting that the same result is obtained even for entangled coherent state light. This reveals that squeezing and optimization of the squeezing parameter $r_e$ are necessary in order to make the $N$-dependence be a Heisenberg limit. 

\section{Obtainable image resolution}

In the following, we will perform an estimate of imaging resolution. We show that the obtainable image resolution is a direct consequence of the integer $Q$, which is the highest-order index of the prolate spheroidal wave functions that are used to reconstruct the object field in the previous section. Recall that the minimum image noise from the MEMSL imaging technique is given in Eq. \eqref{eq:MEMSL-noise}, which is inversely proportional to $\lambda_Q$. We note that, if the index $Q$ increases by one, $\lambda_Q$ decreases significantly (orders of magnitude for a sufficiently large $Q$), which demands to limit $Q$. To determine $Q$, we cut the highest order index $Q$ where the accumulated imaging noise with the summation of the prolate spheroidal functions up to $(Q-1)$st order equals the total number of photons used (i.e., the signal-to-noise ratio is unity). It is due to the observation that higher $Q$ will suffer more from measurement noise since higher-order prolate spheroidal functions have faster spatial oscillations, causing significantly sharper object-to-image attenuation (hence requiring greater amplification when reconstructing the object from the image, which also increases the amplification of the noise). We also exploit the fact that the noise is zero for for odd spheroidal functions (see the definition of $A_j$ in equation \eqref{eq:Aj} and the parity of the odd orders of prolate spheroidal functions (Appendix \ref{sec: Prolate Spheroidal Functions and Eigenvalues})).  As a result, by using an odd value of $Q$ as the cut-off, we can obtain a resolution corresponding to that odd value of $Q$ while the noise corresponds only to the even $(Q-1)$st-order spheroidal function. Therefore, we limit $Q$ such that
\begin{equation}
	\sigma_{Q-1}^2 \approx M N, \label{eq:sigma-Q}
\end{equation}
where we used the fact that the noise energy equals $\sigma^2$. Then, we obtain the condition to determine $Q$ in MEMSL imaging such that
\begin{equation}
	\frac{A_{Q-1}^2}{\lambda_{Q-1}^2} \lessapprox  8 M^2 N^2 (1 + M N). \label{eq:Q}
\end{equation} 
One chooses the maximum integer $Q$ that satisfies this inequality. 

Once the value of $Q$ is determined using the above, the obtainable resolution $D$ is determined by considering the fact that the prolate spheroidal function of order $Q$ ($\psi_Q (s')$) has approximately $Q$ number of zeros in the interval $[-1,1]$ \cite{slepian1965eigenvalues} and $Q+1$ number of mountains and valleys in the same interval (see Fig. \ref{fig:splenian_functions}). Hence, the image resolution is given as
\begin{equation}
	D \simeq \frac{Y}{Q+1} = R \frac{S}{Q+1}, \label{eq:resolution}
\end{equation}
where $S = 2 c/ \pi = d Y/\lambda f$ is the Shannon number, and $R = Y/S = \lambda f/d$ is the Rayleigh resolution length. 

\section{Impact of optical losses}

Optical losses degrade both the noise-squeezing and the entanglement in MEMSL. It is because optical losses couple the probing optical fields used in the imaging to the environments, leading to leakage of information in an irreversible way. Therefore, optical losses will degrade the imaging performance. In this section, we analyze the impact of optical losses on the imaging noise. 

In Guo \textit{et al.}, the loss is modeled as a lumped single beam splitter located between the light source and the sample \cite{guo2020distributed}, which models loss mechanisms in various locations into a single lumped loss. In modeling point of view, this approach is much more attractive than other alternatives because it allows a tractable treatment of the optical loss. The position of the lumped loss element is not important as one can scale the model beam splitter's ratio appropriately according to the actual loss amount. We follow this convention, modeling the optical loss as the probing light undergoing a beam splitting action:
\begin{equation}
	b^{(m)} \rightarrow \sqrt{\tau} b^{(m)} + \sqrt{1 - \tau} d^{(m)}_v,
\end{equation}
where $d^{(m)}_v$ is the annihilation operator of the external vacuum mode that is introduced via the beam splitter. Here, $\tau$ is the beam splitter's splitting ratio, which models the optical loss of $1-\tau$. We note that $\langle d^{(m)}_{v} \rangle = 0$ and $\langle (\Delta d^{(m)}_{v1,v2})^2 \rangle = 1/4$ where $d^{(m)}_{v1} = (d^{(m)}_v + d^{(m)\dagger}_v)/2$ and $d^{(m)}_{v2} = (d^{(m)}_v - d^{(m)\dagger}_v)/(2i)$.

With this, we proceed to modify the previous analysis, now incorporating the optical loss. First, we study the impact of optical loss on MEMSL imaging: we modify the recipe Eq. \eqref{eq:a-MEMSL} into the following, after reflecting the fact that only $\sqrt{\tau}$ ratio of the probe light will arrive at the object after the loss:
\begin{equation}
	\tilde{a}^\mathrm{avg} (s') = \frac{\sqrt{\tau}}{M} \sum_{m=1}^M b^{(m)} e^{i \tilde{\phi}(s')} = \sqrt{\frac{\tau}{M}} b_e e^{i \tilde{\phi}(s')}.
\end{equation}
This leads to
\begin{align}
	\tilde{a}_2^\mathrm{avg} (s') &\approx \alpha_e \sqrt{\frac{2 \tau}{M}} \tilde{\phi}(s'), \nonumber \\
	\tilde{\phi} (s') &= \frac{1}{\alpha_e \sqrt{2 \tau M}} \sum_{j=0}^Q \frac{\left( \sum_{m=1}^M e^{(m)}_{j2} \right)}{\lambda_j} \psi_j (s').
\end{align}
In addition, Eq. \eqref{eq:ej2} is modified to
\begin{align}
	e^{(m)}_{j2} =& \left(\int_{-\infty}^\infty ds' \psi_j (s') \right) \left( \sqrt{\tau} b^{(m)}_2 + \sqrt{1 - \tau} d^{(m)}_{v2} \right) \nonumber \\
	&+ \left( \int_{-1}^1 ds' \phi (s') \psi_j (s') \right) \left( \sqrt{\tau} b^{(m)}_1 + \sqrt{1 - \tau} d^{(m)}_{v1} \right).
\end{align}
Therefore, we calculate
\begin{align}
	&\int_{-1}^1 ds' \langle (\Delta \tilde{\phi} (s'))^2 \rangle  \nonumber \\
	&= \frac{1}{2 \tau \alpha_e^2 M} \sum_{j=0}^Q \frac{\langle \Delta (\sum_{m=1}^M e^{(m)}_{j2})^2 \rangle}{\lambda_j^2} \nonumber \\
	&\approx \frac{1}{2 \tau \alpha_e^2 M} \times \nonumber \\
	& \sum_{j=0}^Q \frac{ A_j^2 \left( M \tau \langle (\Delta b_{e2})^2 \rangle + (1 - \tau) \sum_{m=1}^M \langle (\Delta d^{(m)})^2 \rangle \right)}{\lambda_j^2} \nonumber \\
	&= \frac{e^{-2 r_e} + 1/\tau - 1}{8 \alpha_e^2} \sum_{j=0}^Q \frac{A_j^2}{\lambda_j^2} \nonumber \\
	& \approx \frac{A_Q^2}{8 \alpha_e^2 \lambda_Q^2} \left( e^{-2 r_e} + 1/\tau - 1 \right).
\end{align}
Hence, we obtain a modified imaging error of lossy MEMSL imaging:
\begin{equation}
	\sigma \approx \frac{A_Q}{2 \sqrt{2} \alpha_e \lambda_Q} \sqrt{e^{-2 r_e} + 1/\tau - 1}.
\end{equation}
We note that the above reduces to the lossless MEMSl imaging error in Eq. \eqref{eq:lossless-error} as $\tau \rightarrow 1$ (i.e., lossless). We note that this is a monotonically decreasing function of $\tau$, implying that the imaging error reduces monotonically as $\tau$ increases (i.e., the loss becomes smaller). 

As the loss increases, the number of photons in each mode on the sample reduces. If we are interested in setting the same number of photons $N$ on the sample regardless of the loss, we will have to increase either the displacement $\alpha_e$ or the squeezing parameter $r_e$ as the number of photons is given as Eq. \eqref{eq:photon-number}. It is possible to re-optimize the system parameters by re-designing the squeezing parameter $r_e$ and the displacement $\alpha_e$ to minimize the imaging error for given optical loss $1-\tau$ so that one can still have the same number of photons $N$ impinging on the sample, but now with much higher initial optical power. If one designs the system to have the same $N$, it is intuitively expected that the imaging error may relatively stay to be similar. 

With the optical loss, we modify the number of photons in each branch of MEMSL since only $\tau$ ratio of photons will arrive at the object after the loss:
\begin{equation}
	N_e = \frac{\tau}{M} (\alpha_e^2 + \sinh^2 r_e).
\end{equation}
We then optimize $\alpha_e$ and $r_e$ while fixing $N_e = N$ as we discussed above, yielding the following optimal solutions that minimize the noise $\sigma$:
\begin{align}
	&e^{-r_e} = \sqrt{\frac{\Lambda + \tau}{1 + 4 M N + \tau}}, \nonumber \\
	 &\alpha_e = \nonumber \\
	 &\frac{1}{\sqrt{2}} \sqrt{\frac{8 M^2 N^2 (1 - \tau) - \tau(\Lambda - 1) - 2 MN (\tau(\Lambda + 2 \tau -2)-1)}{(1 - \tau) \tau (1 + 4 MN + \tau)}},
\end{align}
where $\Lambda = \sqrt{1 + 4 M N (1 - \tau)}$. Therefore, the optimal (minimal) imaging error is given as
\begin{widetext}
\begin{equation}
	\sigma^\mathrm{opt} = \frac{A_Q}{2 \lambda_Q}\sqrt{\frac{(1 - \tau)\left( 1 + \tau \Lambda \right)}{8 M^2 N^2 (1 - \tau) + 2 M N (1 - \tau (\Lambda + 2 (\tau-1))) + \tau (1 - \Lambda)}}.
\end{equation}
\end{widetext}
It is easy to verify that the above approaches the result of a lossless case in Eq. \eqref{eq:MEMSL-noise} as $\tau \rightarrow 1$. 

Next, we study the impact of optical loss in non-entangled squeezed light imaging with $M$-time repeating measurements. The analysis is similar, but having a crucial difference so that we now have the imaging error
\begin{equation}
	\sigma \approx \frac{A_Q}{2 \sqrt{2 M} \alpha_e' \lambda_Q} \sqrt{ e^{-2 r_e'} + 1/\tau - 1}.
\end{equation}
The optimal solutions are
\begin{align}
	&e^{-r_e'} = \sqrt{\frac{\Lambda'+\tau}{1 + 4 N + \tau}} , \nonumber \\
	&\alpha_e' = \nonumber \\
	&\frac{1}{\sqrt{2}} \sqrt{\frac{8 N^2 (1 - \tau) - \tau (\Lambda'-1) - 2N (\tau(\Lambda' + 2 \tau - 2) - 1)}{(1-\tau)\tau(1+4N + \tau)}},
\end{align}
where $\Lambda' = \sqrt{1 + 4 N (1 - \tau)}$. These lead to the optimal imaging error of
\begin{widetext}
	\begin{equation}
		\sigma^\mathrm{opt} = \frac{A_Q}{2 \lambda_Q\sqrt{M}}\sqrt{\frac{(1 - \tau)\left( 1 + \tau \Lambda' \right)}{8 N^2 (1 - \tau) + 2 N (1 - \tau (\Lambda' + 2 (\tau-1))) + \tau (1 - \Lambda')}}.
	\end{equation}
\end{widetext}
It is easy to verify that this solution approaches that in Eq. \eqref{eq:ne-squeezing} as $\tau \rightarrow 1$ (i.e., lossless). 

Finally, for the case of classical coherent light imaging with $M$-time repeating measurement, the number of photons used requires $\tau {\alpha'_e}^2 = N$. Hence, with a reduced $\tau$, one must increase $\alpha'_e$ to maintain the same number of photons impinging on the sample. However, the imaging error does not depend on $\tau$ because the number of photons measured are maintained to be identical regardless of $\tau$. In this case, the imaging error in Eq. \eqref{eq:sigma-coherent-light} remains the same regardless of optical loss.

\section{Numerical examples and analysis}

We now present numerical examples of our quantum super-resolution scheme. For numerical evaluations of prolate spheroidal functions and their eigenvalues, we use the following \cite{moore2004prolate}:
\begin{align}
	\lambda_n (c) &= \frac{2c}{\pi} R_{0n}^2 (c,1), \nonumber \\
	\psi_n (c,x) &= \sqrt{\frac{\lambda_n (c)}{\int_{-1}^1 dx' S^2_{0n} (c,x')}} S_{0n} (c,x),
\end{align}
where $R_{mn} (c,x), S_{mn} (c,x)$ are the radial and the angular solution with order $m,n$ of the Helmholtz wave equation of the first kind, respectively. We explicitly made $c$ as a parameter of the functions indicating that both the eigenvalues $\lambda_n$ and the prolate spheroidal functions $\psi_n$ are functions of $c$. The following relation is extremely useful to numerically evaluate $A_j$ in Eq. \eqref{eq:Aj} \cite{moore2004prolate} :
\begin{equation}
	\int_{-\infty}^\infty dt \psi_n (c,t) e^{i \omega t} = i^n \sqrt{\frac{2\pi}{c}} \psi_n \left(c, \frac{\omega}{c} \right), \label{eq:infinite-integral}
\end{equation}
which leads to
\begin{equation}
	A_j = \left| \int_{-\infty}^\infty ds \psi_j (c,s) \right| = \sqrt{\frac{2 \pi}{c}} \left| \psi_j (c,0) \right|.
\end{equation}
For calculating the values of $\psi_n (c,x)$, we used the built-in functions of a python package (Scipy), namely, $R_{mn} (c,x) =$ \texttt{scipy.special.pro\_rad1}(m,n,c,x) and $S_{mn} (c,x) =$ \texttt{scipy.special.pro\_ang1}(m,n,c,x), respectively. These numerical packages are able to evaluate the cases for $|x|<1$. For the numerical evaluation for $|x|\ge 1$, we utilize the Eq. \eqref{eq:xl1}:
\begin{equation}
	\psi_j (s) = \frac{1}{\lambda_j} \int_{-1}^1 ds' \, \psi_j (s') \frac{\sin [c(s'-s)]}{\pi(s'-s)}.
\end{equation}

We first demonstrate the utility of the MEMSL imaging technique where we assume lenses with a focal length $f = 10$ mm, the wavelength of light $\lambda = 780$ nm, and a lens diameter $d$ of 2 inches. In this case, the Rayleigh diffraction limit is $\lambda f / d = 154$ nm. We also assume that the object has a size $Y = 300$ nm. These parameter values imply $c = \pi d Y/2 \lambda f = 3.07$. We then assume that the MEMSL has eight entangled modes ($M=8$), and each mode has 6 photons ($N=6$). In this case, the optimal squeezing and displacement of the input single-mode squeezed mode to MEMSL generator from Eq. \eqref{eq:optimal_solution_MEMSL} are $e^{- r_e} = 0.10$ and $\alpha_e = 4.92$.  Also, this example implies that the highest order number $Q$ according to the Eq. \eqref{eq:Q} is 7 (since the noise energy corresponding to $Q-1 = 6$ is on the same order as the total photon number), implying that the resolution one can achieve is 38 nm according to Eq. \eqref{eq:resolution}. This example of MEMSL imaging beats the Rayleigh diffraction limit deeply by roughly a factor of 5 with such a small number of photons used (total 48 photons). On the other hand, in order to accomplish the same level of imaging error, the classical imaging using non-entangled non-squeezed coherent state light with $M$-time repeating measurement would require 294 photons each time for averaging over eight times (total 2352 photons), which is roughly two orders-of-magnitude larger number of photons on the sample than MEMSL imaging. Therefore, we learn that MEMSL imaging significantly reduces the required number of photons.

In order to simulate the imaging system with quantum noise, we utilize the Eq. \eqref{eq:ej2}, which is refined using Eq. \eqref{eq:infinite-integral}:
\begin{equation}
	e^{(m)}_{j2}  = b_2^{(m)} \sqrt{\frac{2 \pi}{c}} \psi_j (c,0) + b_1^{(m)} \int_{-1}^1 ds' \phi (s') \psi_j (s') . \label{eq:eb21}
\end{equation}
The quantum noise enters via $b_2^{(m)}$ and $b_1^{(m)}$, which are quadrature operators. A more informative quantity is the average $(1/M) \sum_{m=1}^M e_{j2}^{(m)}$ since this quantity is directly used to reconstruct $\tilde{\phi} (s')$, the original object's spatial phase distribution. For the case of MEMSL imaging, we use the Eq. \eqref{eq:eb}, where the noise is given in Eq. \eqref{eq:be12_stat} with Gaussian noise statistics. For coherent light imaging method, we use Gaussian noise with $\langle b_{1}^{(m)} \rangle = \sqrt{2} \alpha'$, $\langle b_{2}^{(m)} \rangle = 0$, $\langle (\Delta b_{1,2})^2 \rangle = 1/4$. The image field's second quadrature is given as $e_2^{(m)}(s') = \sum_{j=1}^\infty e_{j2}^{(m)} \psi_j (s')$, which now contains the appropriate quantum noise for different cases of MEMSL imaging or non-entangled squeezed/coherent light imaging with $M$-time repeating measurement.

\begin{figure}
	\centering
	\includegraphics[width=0.9\linewidth]{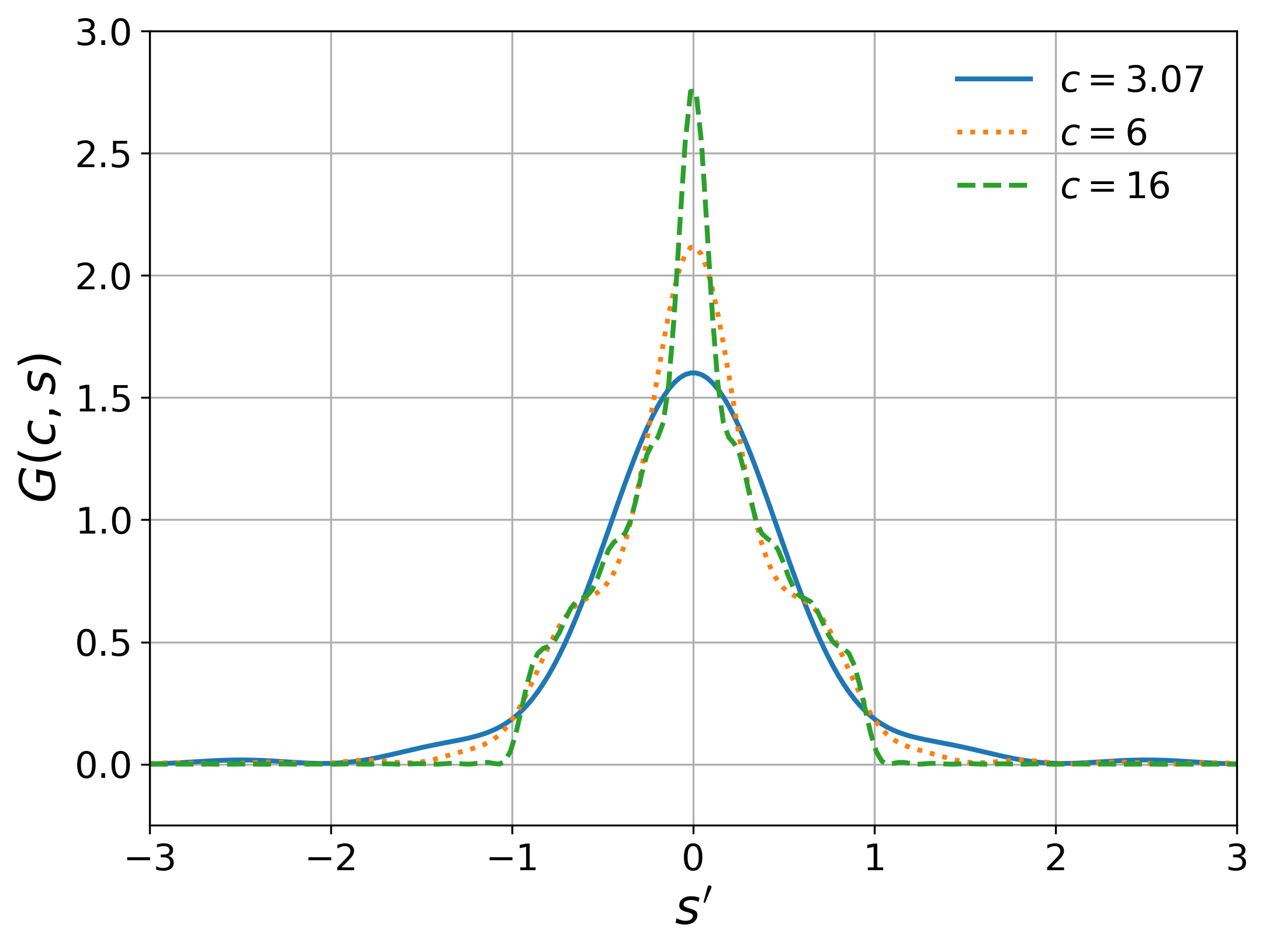}
	\caption{Numerical evaluation of $G(c,s')$ for various $c$ values.} \label{fig:G}
\end{figure}

An alternative numerical simulation approach is to directly generate random data for $ e_2^\mathrm{avg} (s') = (1/M)\sum_{m=1}^M e_2^{(m)} (s')$ for each point $s'$. Note that, from Eq. \eqref{eq:e-a-transform}, we obtain the average field:
\begin{equation}
	\langle e_2^\mathrm{avg} (s') \rangle = \int_{-1}^1 ds' \frac{\sin[c(s'-s)]}{\pi(s'-s)} \left( \frac{1}{M} \sum_{m=1}^M \langle a_2^{(m)} (s') \rangle \right). \label{eq:ej2m}
\end{equation}
With $\langle a_2^{(m)} (s') \rangle \simeq \langle b_2^\mathrm{(m)} \rangle \phi(s')$ (see Eq. \eqref{eq:a-b}), the above average is easily calculated. For the calculation of the variance $\langle (\Delta e_2^\mathrm{avg} (s'))^2 \rangle$, we separate the case of MEMSL imaging and the case of the coherent light imaging repeating $M$-times. For MEMSL imaging, the calculation is
\begin{align}
	&\langle (\Delta e_2^\mathrm{avg} (s'))^2 \rangle  \nonumber \\
	&= \frac{1}{M^2} \times \nonumber \\
	& \sum_{m=1}^M\sum_{j,j'=0}^\infty \left( \langle e_{2j}^{(m)} e_{2j'}^{(m)} \rangle - \langle e_{2j}^{(m)} \rangle \langle e_{2j'}^{(m)} \rangle \right) \psi_j (s') \psi_{j'} (s') \nonumber \\
	&= \frac{1}{M^2} \sum_{m=1}^M \sum_{j=0}^\infty \langle (\Delta e_{2j}^{(m)})^2 \rangle \psi_j^2 (c,s') \nonumber \\
	&\simeq \frac{\langle (\Delta b_{2e})^2 \rangle}{M} \sum_{j=0}^\infty  \frac{2 \pi}{c} \psi_j^2 (c,0) \psi_j^2 (c,s') \nonumber \\
	&= \frac{G(c,s')}{4M(1+2MN)} , \label{eq:MEMSL-point-error}
\end{align}
where
\begin{equation}
	G(c,s') = \sum_{j=0}^\infty  \frac{2 \pi}{c} \psi_{2j}^2 (c,0) \psi_{2j}^2 (c,s').
\end{equation}
Here, the second equation in \eqref{eq:MEMSL-point-error} used the fact that $e_{2j}$ and $e_{2j'}$ are independent random variables unless $j=j'$. The third equation used the MEMSL relation in \eqref{eq:MEMSL-eq} and the fact that the noise contribution of the second term in Eq. \eqref{eq:eb21} is negligible. The last line used the optimal solution for MEMSL in Eq. \eqref{eq:optimal_solution_MEMSL} and the fact that $\psi_j (c,0) = 0$ for odd $j$s (see Fig. \ref{fig:splenian_functions} (b)). It is known that the following sum converges (see Eq. (78) of Moore and Cada \cite{moore2004prolate}):
\begin{equation}
	\sum_{j=0}^\infty \frac{2\pi}{c} \psi_{2j}^2 (c,0) = E^c,
\end{equation}
where $E^c$ is a constant. Hence, $G(c,s')$ also converges because $\psi_{2j}^2 (c,s') \leq \psi_{c,0}^2 (0)$ for all $j\ge 1$. In fact, the summation quickly converges numerically (the numerical accuracy is sufficient even if we sum only up to $j = 20$). Numerical evaluations of the function $G(c,s')$ are shown in Fig. \ref{fig:G}. Higher $c$ values concentrate more towards the center. Incidentally, the domain of $s'$ where $G(c,s')$ is large coincides with region of dense information. Since both the expected value and the variance of the Gaussian random variable $e_2^\mathrm{avg} (s')$ are known for each $s'$, we can numerically simulate the entire noisy $e_2^\mathrm{avg} (s')$ for all $s'$. 

For coherent light imaging repeating $M$ times, the following holds:
\begin{align}
	&\langle (\Delta e_2^\mathrm{avg} (s'))^2 \rangle \nonumber \\
	&\simeq \frac{1}{M^2} \sum_{m=1}^M \sum_{j=0}^\infty \langle (\Delta b_{2}^{(m)})^2 \rangle \frac{2 \pi}{c} \psi_j^2 (c,0) \psi_j^2(c,s') \nonumber \\
	&= \frac{G(c,s')}{4M}. \label{eq:coh-point-error}
\end{align}
Likewise, because we know both the expected value and the variance of $e_2^\mathrm{avg} (s')$ for all points $s'$ for the classical imaging using coherent state light, we can numerically simulate the entire noisy $e_2^\mathrm{avg} (s')$ for all $s'$.

Let us compare the above two results in Eq. \eqref{eq:MEMSL-point-error} and \eqref{eq:coh-point-error}: MEMSL imaging significantly reduces the error in each measurement data $e_2^\mathrm{avg} (s')$ at all points $s'$ because the noise variance is inversely proportional to roughly $M^2$, which is due to the entanglement in the MEMSL branch modes, and also to the number of photons $N$ in each MEMSL branch mode. In contrast, the coherent light imaging lacks this feature, and the error in $e_2^\mathrm{avg} (s')$ is inversely proportional to only $M$, due to lack of entanglement in the independent $M$-time measurement. It is remarkable to observe that the number of photons $N$ does not reduce the error of $e_2^\mathrm{avg}(s')$ at each individual point $s'$ for the classical imaging using coherent state light, while it reduces the accumulated error $\sigma$ (the phase distribution error, see Eq. \eqref{eq:sigma-coherent-light}). This can be understood because the error in phase distribution estimate follows $\Delta \tilde{\phi}(s) \propto \Delta a^\mathrm{avg} (s) / \sqrt{N}$ (c.f., Eq. \eqref{eq:a_avg_phi}). So, while the phase distribution estimate error reduces as $1/\sqrt{N}$ in the coherent light imaging repeating $M$-times, the measured value $a^\mathrm{avg} (s)$, which is obtained from $e^\mathrm{avg} (s')$, is not affected by $N$. 

\begin{figure}
	\centering
	\begin{subfigure}[b]{1\linewidth}
		\includegraphics[width=0.9\linewidth]{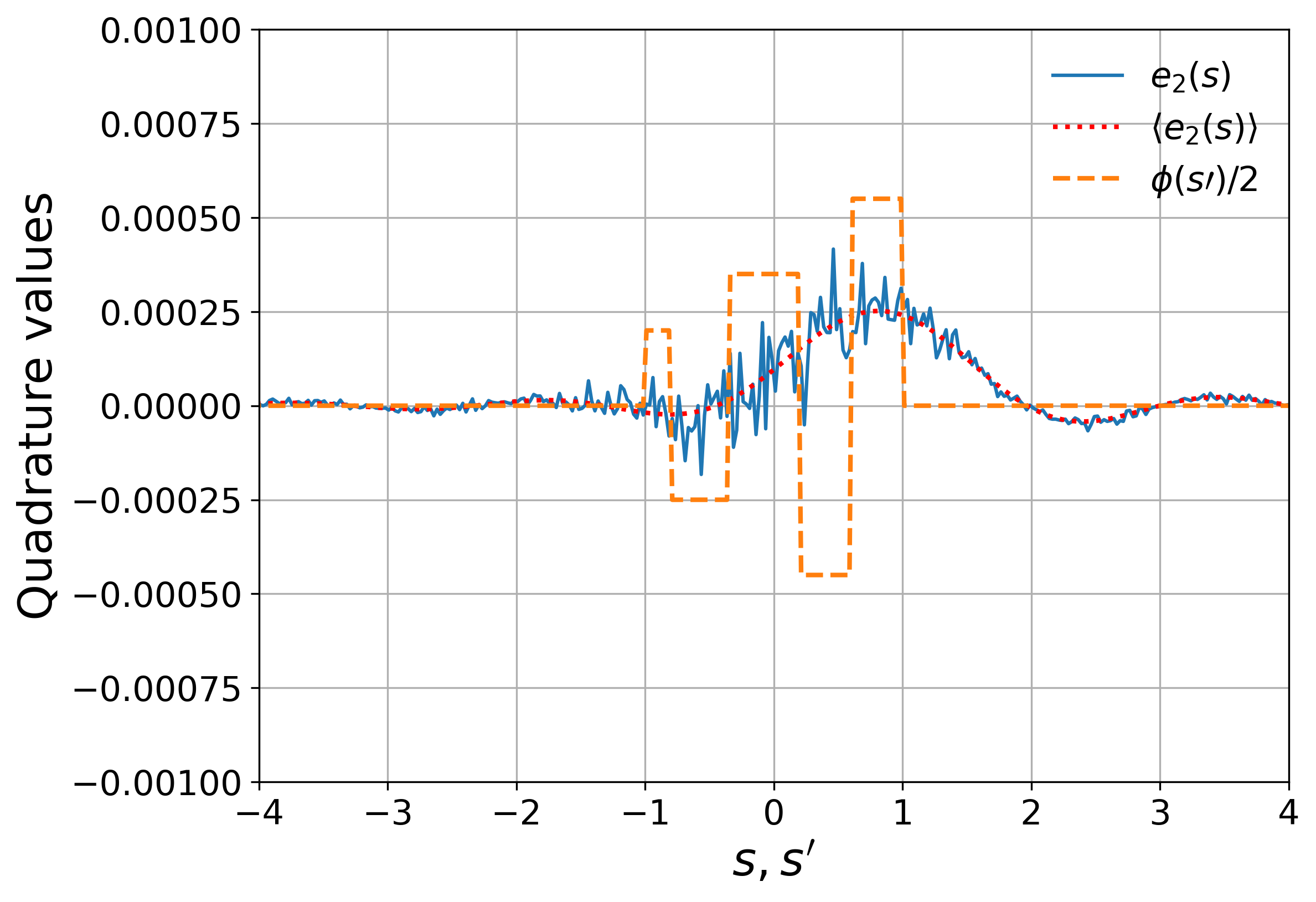}
		\caption{}
	\end{subfigure}
	\begin{subfigure}[b]{1\linewidth}
		\includegraphics[width=0.9\linewidth]{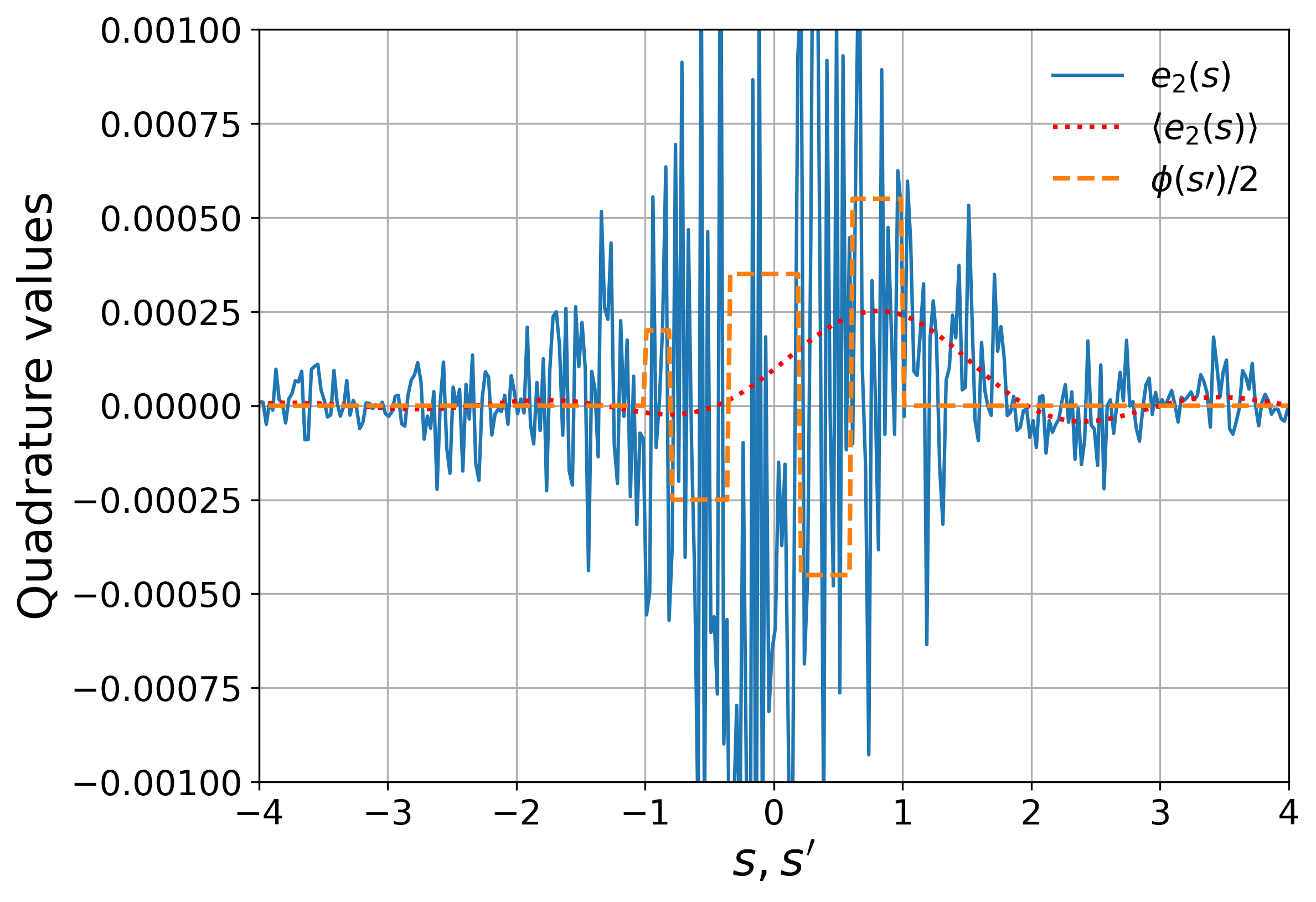}
		\caption{}
	\end{subfigure}
	\caption{Image field quadrature values (solid lines) with added quantum noise resulting from the object's phase distribution (dashed lines). The measurement is repeated for $50000$ times with $M = 4$ and $N = 12$ for both (a) MEMSL imaging and (b) coherent light imaging. The imaging system's space-bandwidth is $c = 3.07$.}\label{fig:e2}
\end{figure}

We note that achieving high-level squeezing is hard. For example, the best \textit{measured} noise squeezing result so far is 15 dB \cite{vahlbruch2016detection}. In imaging, the actual measured level of squeezing matters because only imaging performance depends on measured data. According to the optimal squeezing solution in Eq. \eqref{eq:optimal_solution_MEMSL}, the optimal squeezed noise power ($e^{-2 r_e}$) is desired to be $1/(1+2MN)$. The 15 dB squeezing implies only, for example, four photons ($N = 4$) in each mode for eight modes ($M = 8$). The hardship in accomplishing large squeezing forces averaging over multiple independent measurements, rather than accomplishing an excessively large level of squeezing. In contrast, the coherent light imaging repeating $M$-times does not see any difference in the roles of $M$ and $N$ (see Eq. \eqref{eq:sigma-coherent-light}), which makes sense physically since repeating more times using coherent light is exactly same as adding more photons in a single measurement as the independent photons reduce the noise as $\sim 1/\sqrt{N}$. For numerical simulations, we assume a fictitious one-dimensional object plane's phase distribution as shown in Fig. \ref{fig:e2} (dashed-line), which is designed intentionally to be asymmetric to accommodate both the even and the odd orders of prolate spheroidal functions. Then, we performed numerical simulation of the image plane fields for both MEMSL imaging with $M=8$ and $N=6$ (corresponding to 20-dB-squeezed input light (i.e., $10 \log_{10} (e^{-2 r_e}) = -20$) to the array of balanced beam splitters to generate the MEMSL light)  (which used the same $M=8$ and $N=6$), both averaging over 50000 independent measurements. Fig. \ref{fig:e2} shows numerically simulated image plane quadrature distribution $e_2(s)$ with added quantum noise for MEMSL imaging and coherent light imaging. It is clearly shown that the MEMSL imaging's (simulated) measured field data shows significantly reduced noise compared to the coherent light imaging.

\begin{figure}
	\centering
	\includegraphics[width=\linewidth]{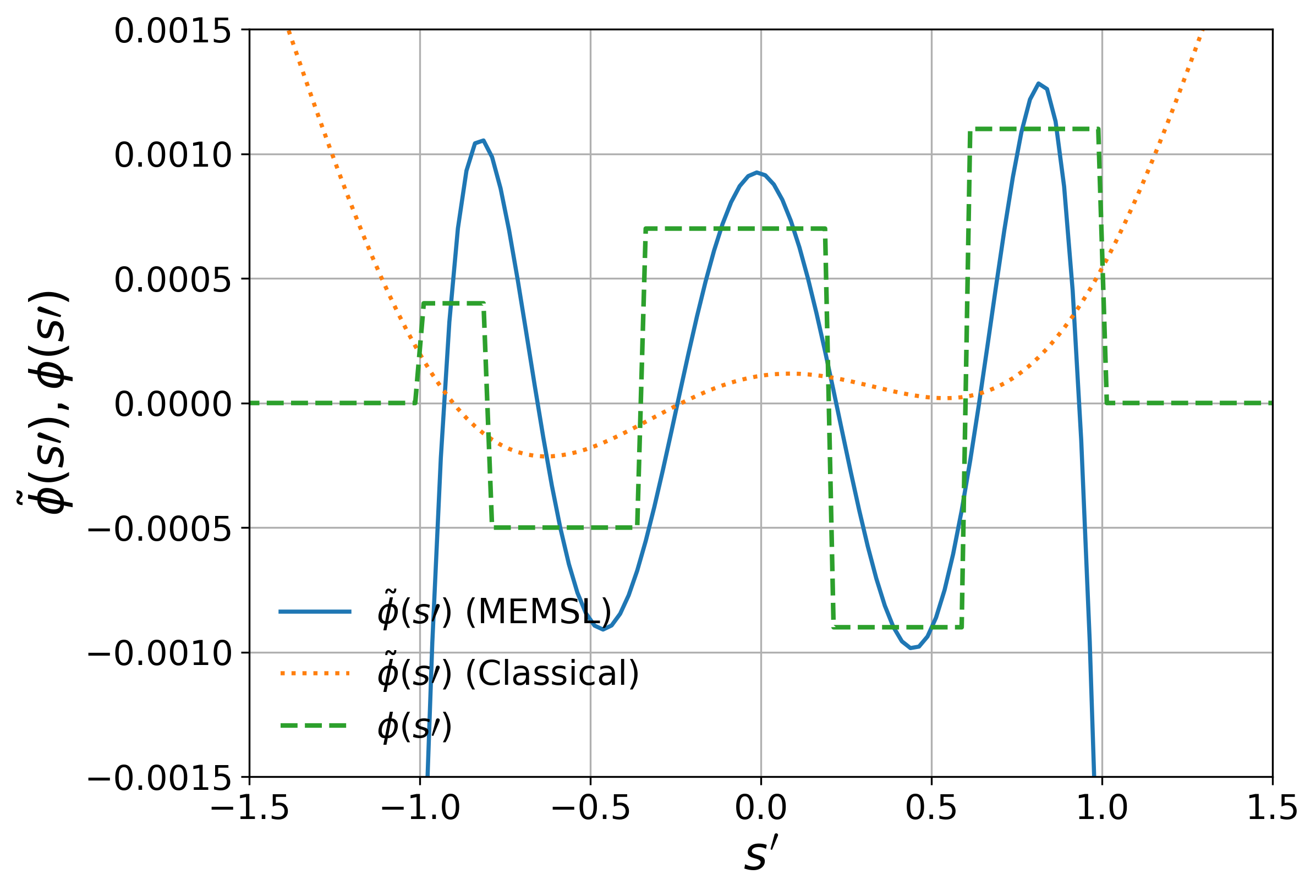}
	\caption{Reconstructed object plane's optical phase distributions $\tilde{\phi}(s')$ from MEMSL imaging (solid line) and classical light imaging (dotted line), along with the original accurate test object plane's optical phase distribution $\phi(s')$ (dashed line). } \label{fig:phi_est}
\end{figure}

We follow the reconstruction recipe in Eq. \eqref{eq:recon-recipe} for estimating $\tilde{\phi}(s')$. For the maximum index $Q$, we follow the criterion in Eq. \eqref{eq:sigma-Q} for both MEMSL and classical imaging, which are good for a single-shot object field estimate. For averaging  $N^\mathrm{avg}$-time independent repeating measurements, the equation must slightly change to $\sigma \rightarrow \sigma/\sqrt{N^\mathrm{avg}}$ as the repeating measurements are independent (i.e., unentangled). Also, the total number of photons used must change $MN \rightarrow MN N^\mathrm{avg}$. These change the maximum $Q$ criterion to be
\begin{equation}
	\frac{A_{Q-1}^2}{\lambda_{Q-1}^2} \leq 8 M^2 N^2 N^\mathrm{avg 2} (1 + MN).
\end{equation}
According to this, the estimated maximum $Q$ is 7 for MEMSL. On the other hand, for the coherent light imaging repeating $M$-times, we use the Eq. \eqref{eq:sigma-coherent-light} for applying to Eq. \eqref{eq:sigma-Q}
\begin{equation}
		\frac{A_{Q-1}^2}{\lambda_{Q-1}^2} \leq 8 M^2 N^2 N^\mathrm{avg 2},
\end{equation}
which produces the estimated maximum $Q = 5$. We found that adding a higher-order prolate spheroidal functions made the estimate $\tilde{\phi}(s')$ quickly diverging due to the denominator $\sqrt{\lambda_j}$ in Eq. \eqref{eq:tilde-ae}, which decreases fast by multiple order-of-magnitudes with a slightly increased index $j$. Fig. \ref{fig:phi_est} shows the reconstructed object plane's optical phase distribution $\tilde{\phi} (s')$ for both MEMSL imaging and coherent light imaging. While the reconstructed optical phase distribution from MEMSL imaging accurately shows three lobes with somewhat inaccurate heights, that of the classical light imaging completely missed the salient features of the object's optical phase distribution (i.e., wrong number of lobes, completely wrong heights). 

We note that our numerical example demonstrated the efficacy of our MEMSL quantum imaging for probing nanoscale structures: revealing such a complex and fine structure with size far below the Rayleigh diffraction limit demands the imaging performance to beat the Rayleigh diffraction limit deeply. The accomplished imaging resolution from MEMSL imaging in the numerical example is roughly 38 nm ($=300/(7+1)$ with $Q=7$) whereas the Rayleigh diffraction limit in this case was 154 nm. We note that the obtainable resolution can be improved by several means. Having more entangled modes ($M$) with increased number of photons ($N$) improves the optical resolution at the cost of large squeezing in the input single-mode squeezed light for MEMSL generator. In practice, increasing the number of independent measurement trials (i.e., increasing $N^\mathrm{avg}$) is a viable solution. 

We emphasize that, with the same number of modes $(M)$ and the same number of photons $(N)$ for a single measurement, MEMSL imaging is always superior to the classical light imaging due to the combined benefit from the squeezing and the entanglement in MEMSL. On the contrary, the effect of increasing $N^\mathrm{avg}$ is basically identical between MEMSL and classical light imaging since the measurements are independent from each other for every measurement trial. 

\begin{figure}
	\centering
	\subfloat[][]{\includegraphics[width=0.90\linewidth]{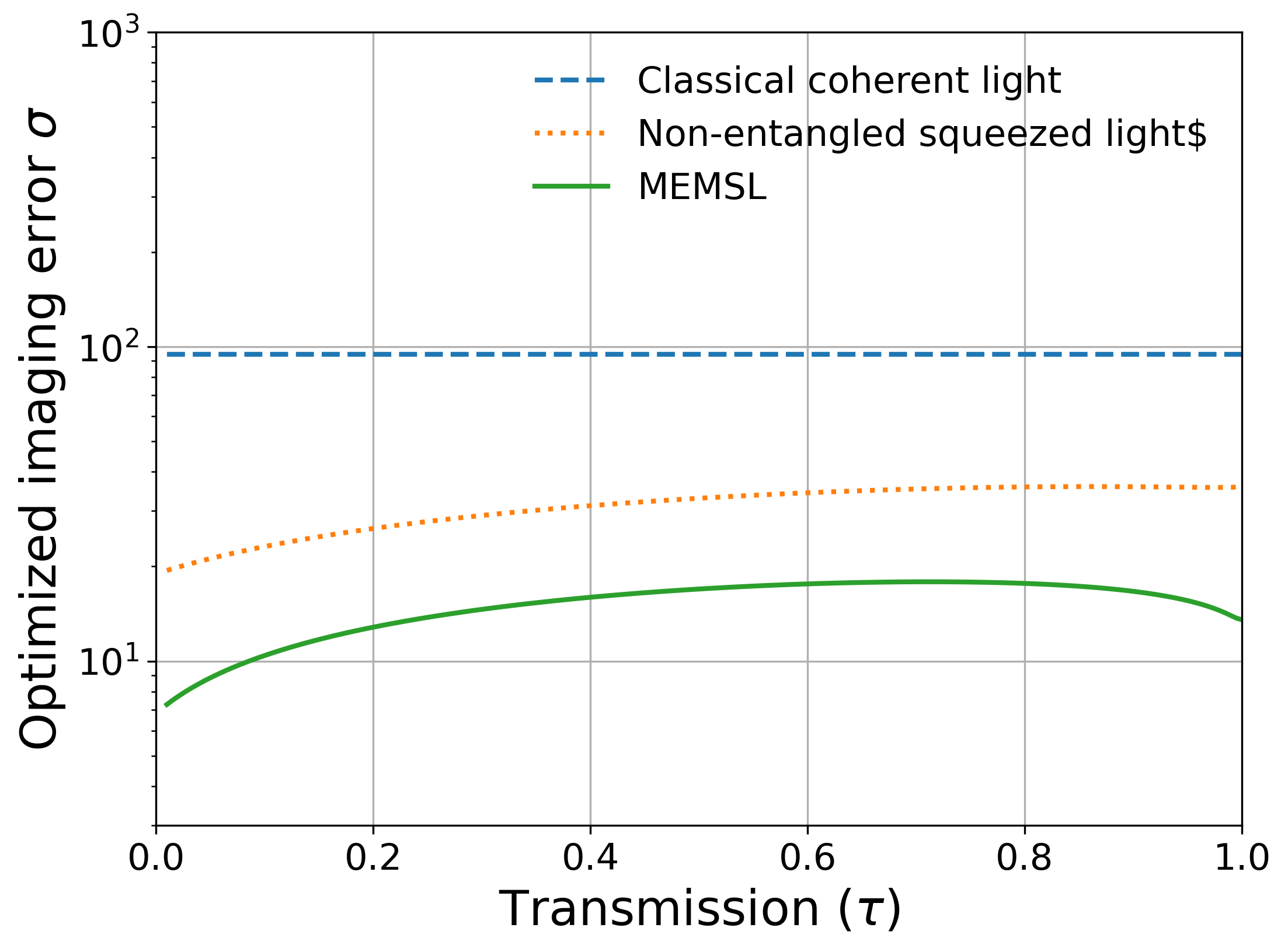}} \\
	\subfloat[][]{\includegraphics[width=0.99\linewidth]{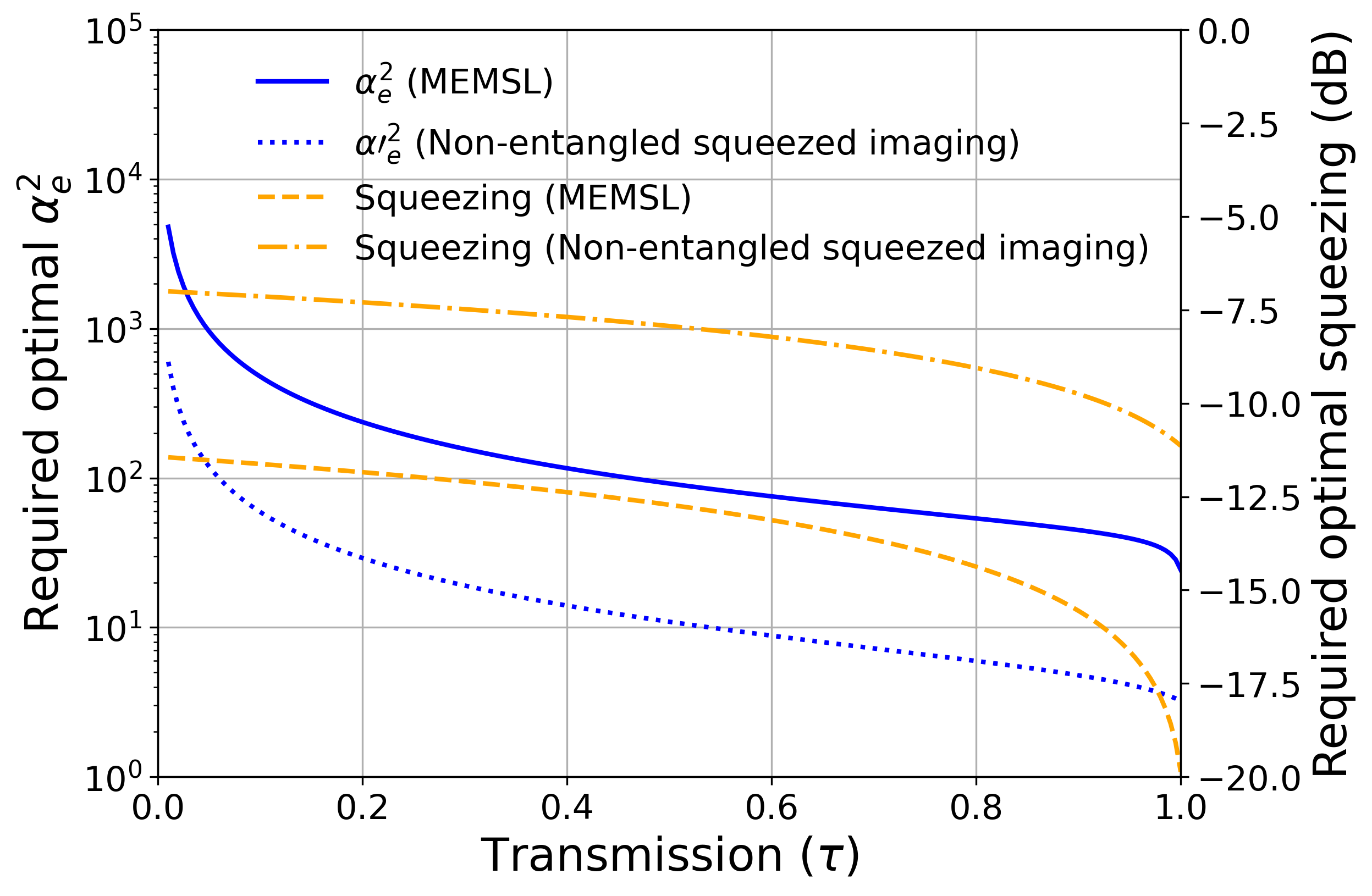}}
	\caption{(a) Comparison of optimized imaging error $\sigma$ from classical coherent light imaging (dashed), non-entangled squeezed light imaging (dotted), and MEMSL imaging (solid) as functions of light transmission $\tau$. (b) Optimal values of displacement $\alpha_e^2$ and initial squeezing ($10 \log_{10} (e^{-2 r_e})$) of MEMSL imaging and non-entangled squeezed light imaging as a function of light transmission $\tau$.} \label{fig:optimized-imaging-error}
\end{figure}

Next, we compare the optimized imaging errors $\sigma$ of MEMSL imaging, non-entangled single-mode squeezed light imaging repeating multiple times, and classical imaging using non-entangled non-squeezed coherent light imaging for various values of $\tau$ (the optical loss is given as ($1-\tau$)) (see Fig. \ref{fig:optimized-imaging-error}). It is clear that the optimized imaging error of MEMSL imaging is approximately an order-of-magnitude smaller than the classical coherent light imaging for various values of $\tau$. The imaging error of MEMSL imaging is also consistently better than that of non-entangled repeating squeezed light imaging as well. While the imaging error values stay relatively similar for different $\tau$ values as expected, it is somewhat counter-intuitive to observe that lower transmission  values (higher optical losses) lead to smaller optimized imaging error for both MEMSL imaging and non-entangled squeezed light imaging repeating multiple times.  

We also compare the optimal configuration of MEMSL imaging and non-entangled repeating squeezed light imaging (Fig. \ref{fig:optimized-imaging-error} (b)). We note that the displacement $\alpha_e$ and the level of squeezing $e^{-2r_e}$ of MEMSL imaging represent the properties of the single-mode squeezed light used as an input for MEMSL generation while $\alpha'_e$ and $e^{-2r'_e}$ of non-entangled squeezed light imaging repeating multiple times are properties of the individual light source for each measurement, which is somewhat similar to the individual mode of MEMSL. Therefore, it is difficult to give a direct fair comparison between $\alpha_e$ and $\alpha'_e$, and also $e^{-2r_e}$ and $e^{-2r'_e}$. With this caveat, we find that $\alpha_e^2$ is nearly the same as $M {\alpha'}_e^2$ for all values of $\tau$, which makes good physical sense, since dividing the single-mode input squeezed light into $M$ modes corresponds to the individual squeezed light in the non-entangled repeating squeezed light imaging. Naturally, the optimal configuration for $alpha_e^2$ and ${\alpha'_e}^2$ increases as $\tau$ decreases (i.e., optical loss increases). In contrast, the optimal configuration for squeezing is reverse, indicating that the role of squeezing becomes less important as the loss increases, and the imaging performance is rapidly compensated by large displacement values $(\alpha_e^2, {\alpha'_e}^2)$, which is in consistence with the expectation that optical loss will quickly degrade the benefit coming from squeezing. It is also important to note that the required squeezing of MEMSL is much more demanding than that of non-entangled squeezed light imaging repeating multiple times as shown clearly in the graph in Fig. \ref{fig:optimized-imaging-error} (b). The reason for this is that the squeezing in the single-mode squeezed input for generating MEMSL is a resource to create the entanglement among the $M$ modes of MEMSL, which will accomplish much reduced imaging error in MEMSL imaging than non-entangled repeating squeezed light imaging. Therefore, the extra degree of squeezing requirement of MEMSL is the price to pay to accomplish a better imaging performance.

\section{Discussions and conclusion} \label{sec:disc}

We presented a new label-free quantum imaging technique using MEMSL as probe light. It is clearly shown that MEMSL's imaging performance is superior compared to other methods presented here, namely, non-entangled single-mode squeezed light imaging repeating multiple times and \textit{classical} imaging using coherent state light repeating multiple times. The MEMSL imaging collects the measurement data for each mode in MEMSL and combines them later to reconstruct the original object information. Due to quantum entanglement and squeezing, the imaging error in MEMSL image accomplishes the Heisenberg limit where the error scales as $1/MN$ ($M$: number of entanglement modes in MEMSL and $N$: number of photons impinging on the sample in each MEMSL mode), while that of non-entangled squeezed light imaging repeating multiple times scales as $1/N\sqrt{M}$, and that of \textit{classical} imaging using coherent state light repeating multiple times scales as the standard limit $1/\sqrt{NM}$. The Heisenberg limit $1/M$ is ascribed to the quantum entanglement among $M$-modes in MEMSL whereas the Heisenberg limit $1/N$ is ascribed to the squeezing. 

The performance of MEMSL imaging in real world is limited by how much squeezing one can accomplish. The achievable squeezing limits the product of $MN$ in the optimal imaging configurations (see Eq. \eqref{eq:optimal_solution_MEMSL}). With the practical squeezing level of $15 - 20$ dB, we demonstrated that the imaging error of MEMSL imaging is still approximately an order of magnitude better than that of using coherent state light repeating multiple times. This reduced imaging noise led to more faithful reconstruction of the object information: MEMSL imaging accurately reconstructed quite complex object features with a resolution that deeply beats the Rayleigh diffraction limit while classical imaging using coherent state light repeating multiple times fails to extract such complex object features, when both imaging methods used the same number of modes ($M$) and the same number of photons ($N$) impinging upon the imaging sample. In the case of MEMSL imaging with $c=3.07$, $M=8$, $N=6$, classical imaging using coherent state light repeating multiple times will need approximately two orders-of-magnitude larger number of photons in order to accomplish the same imaging error as MEMSL imaging.

We also provided a comprehensive optimization strategy when imaging system suffers from optical losses. The price to pay in lossy imaging system is to use a squeezed light input to MEMSL generator with a larger displacement (i.e., higher light power) with somewhat less degree of squeezing than the optimal lossless case. We showed that such an arrangement accomplishes the same image quality under the same limit for the number of photons impinging on the sample as the lossless case. While a lossless imaging system sees equal contributions in the number of photons (i.e., $N_e = \alpha_e^2 + \sinh^2 r_e$) from the displacement ($\alpha_e^2$) and squeezing ($\sinh^2 r_e$) to accomplish a good imaging quality, lossy imaging systems rely more on the displacement ($\alpha_e^2$) than squeezing ($\sinh^2 r_e$) since squeezing quickly degrades with respect to optical losses. 

While we derived the theory based on a one-dimensional imaging using one-dimensional prolate spheroidal (Slepian) functions, we note that it is straightforward to adopt higher-dimension prolate spheroidal functions \cite{slepian1964prolate, shkolnisky2007prolate, yang2002two} to expand our theory to higher-dimensional imaging. Since two-dimensional prolate spheroidal functions also satisfy the crucial integral relation similar to Eq. \eqref{eq:Splenian}, which guarantees that the two-dimensional prolate spheroidal functions are still the eigenfunctions after the two-dimensional integrals, all the theory stands roughly the same, except changing the one-dimensional integrals to two-dimensional integrals. With this caveat, our one-dimensional analysis can be appropriately adapted for a two-dimensional imaging application. Furthermore, it is important to mention that, because we normalized the coordidnates $s,s'$ in the object and the image plane, our theory also can be applied to different two-lens imaging systems that have disparate focal lengths, after appropriately scaling normalized coordinates $s,s'$.

\section*{Acknowledgements}

We thank Eric Chatterjee and Philiip Chrostoski for various valuable discussions.

The work was supported by the Laboratory Directed Research and Development program at Sandia National Laboratories, a multimission laboratory managed and operated by National Technology and Engineering Solutions
of Sandia, LLC., a wholly owned subsidiary of Honeywell International, Inc., for the U.S. Department of
Energy’s National Nuclear Security Administration under Contract No. DE-NA-003525. This paper describes objective technical results
and analysis. Any subjective views or opinions that might be expressed in the paper do not necessarily
represent the views of the U.S. Department of Energy or the United States Government.

\appendix

\section{Prolate Spheroidal Functions and Recovery of Objects} \label{sec:general-imaging}

In this section, we briefly present how one recovers the object from the imaging plane measurements, largely based on the results in \cite{kolobov2000quantum} (see Fig. \ref{fig:imaging}).

Passing through the object, the optical field acquires spatially-dependent phases $\phi(s')$: for example, the $m$-th mode in MEMSL becomes
\begin{equation}
	a^{(m)} (s') = b^{(m)} e^{i \phi(s')}, \quad |s'| \leq 1. \label{eq:a-b}
\end{equation}
The goal of imaging is to obtain $\phi(s')$ accurately, allowing for the reconstruction of the object at every point $s'$. The object-to-image transformation is given as follows (see Appendix~\ref{sec: Derivation of Object-to-Image Transformation} for a derivation):
\begin{equation} \label{eq: object-to-image transformation}
L[\psi](s) = \int_{-\infty}^{\infty} ds' \psi(s') \frac{\sin{(c(s'-s))}}{\pi(s'-s)},
\end{equation}
where $c$ is the previously defined space bandwidth product. Note that this accords with the expression given by Kolobov and Fabre \cite{kolobov2000quantum}. In order to reconstruct the object from the image information, we decompose the object field into basis fields. We start with the \textit{in-pupil} eigenfields $\varphi_j$ of the object-to-image transformation $L$, which satisfy the following relationship for $|s| \leq 1$:
\begin{equation} \label{eq: in-pupil series}
L[\varphi_j](s) = \lambda_j \varphi_j(s).
\end{equation}
In the small-object limit, however, the image size extends well beyond the object size, as previously discussed. For each in-pupil field $\varphi_j$, we define the \textit{out-pupil} field $\chi_j$ to represent the part of the image in the region $|s| > 1$, less a proportionality constant that is a function of $\lambda_j$:
\begin{equation} \label{eq: out-pupil series}
\chi_j(s) = \frac{L[\varphi_j](s)}{\sqrt{\lambda_j (1 - \lambda_j)}}.
\end{equation}
For a classical light source, the object could simply be decomposed into the basis of the in-pupil series $\varphi_j$, since no scattering point exists for $|s'| > 1$. However, due to quantum fluctuation, it is necessary to include a superposition of the out-pupil series $\chi_j$ as well, with the corresponding coefficients featuring an expectation value of zero. Therefore, the object plane field is decomposed as follows:
\begin{equation}
	a^{(m)} (s') = \sum_{j=0}^\infty \left( a_j^{(m)} \varphi_j (s') + c_j^{(m)} \chi_j (s') \right),  -\infty < s' < \infty. \label{eq:amsprime}
\end{equation}
Here, $a_j^{(m)}$ and $c_j^{(m)}$ are superposing weights for $j$th order orthogonal functions $\varphi_j(s')$ and $\chi_j(s')$, respectively. Note that $\langle c_j^{(m)} \rangle = 0$ for all values of $j$. In general, Eq.~\eqref{eq: in-pupil series} is solved by the prolate spheroidal wave functions (a.k.a., Slepian functions) $\psi_j (s')$, which satisfy the following conditions \cite{kolobov2000quantum}:
\begin{align}
&L[\psi](s) = \psi(s), \\
\label{eq:xl1}
&\int_{-1}^{1} ds' \psi_j(s') \frac{\sin{(c(s'-s))}}{\pi(s'-s)} = \lambda_j \psi_j(s).
\end{align}
The orthogonal functions $\varphi_j (s')$ and $\chi_j (s')$ relate to $\psi_j (s')$ as follows:
\begin{align}
	&\varphi_j (s') = \left\{ \begin{array}{ll}
		\frac{\psi_j (s')}{ \sqrt{\lambda_j}},& |s'| \leq 1, \\
		0, & |s'|>1.
		\end{array} \right. \\
	&\chi_j(s') = \left\{ \begin{array}{ll}
		0, & |s'|\le 1, \\
		\frac{\psi_j (s')}{\sqrt{1 - \lambda_j}}, & |s'|>1 
		\end{array} \right. \label{eq:Splenian}
\end{align}
where $c = \pi d Y / 2 \lambda f$ is the space-bandwidth product in Eq. \eqref{eq:spatial-bandwidth}. Here, $\lambda_j$'s are the eigenvalues of $j$th order prolate spheroidal wave functions under the integral defined in Eq. \eqref{eq:xl1} with the property $1 \geq \lambda_0 > \lambda_1 > \cdots > 0$. Appendix~\ref{sec: Prolate Spheroidal Functions and Eigenvalues} shows the eigenvalues $\lambda_j$ and both the odd and the even orders of the prolate spheroidal functions, which are all functions of the value $c$. 

For quantum optics treatment, we elevate the superposing weights $a_j^{(m)}$ and $c_j^{(m)}$ to bosonic annihilation operators, satisfying commutation relations $[a_j^{(m)}, a_k^{ (m) \dagger}] = \delta_{j,k}$ and $[c_j^{(m)}, c_k^{ (m) \dagger}] = \delta_{j,k}$ as is well justified in Kobolov and Fabre \cite{kolobov2000quantum}. Then, $a^{(m)} (s')$ becomes a quantum field operator. We also note that the image plane field is decomposed as
\begin{equation}
	e^{(m)} (s) = \sum_{j=0}^\infty e_j^{(m)} \psi_j (s), \quad -\infty < s < \infty. \label{eq:def-e}
\end{equation}
We note the relation between the object plane field and the image plane field (\cite{kolobov2000quantum}, where the integral domain is not explicitly written after the authors added the quantum treatment including the domain $|s'|>1$, but we make it explicit and now add the outside region in the integral):
\begin{equation}
	e^{(m)} (s) = L[a^{(m)}] (s). \label{eq:e-a-transform}
\end{equation}
The image superposition coefficients (which we elevate to bosonic annihilation operators) thus take the following form in terms of the object coefficients (see the derivation in Appendix~\ref{sec: Calculating the Image's Superposition Coefficients}):
\begin{equation}
	e^{(m)}_j = \sqrt{\lambda_j} a^{(m)}_j + \sqrt{1 - \lambda_j} c^{(m)}_j. \label{eq:e-a}
\end{equation}
One has to estimate the average object field $a^\mathrm{avg} (s') = (1/M) \sum_{m=1}^M a^{(m)} (s')$ after one takes the measurement of image field $e^{(m)} (s)$, for all $m = 1,2, \cdots, M$. The recipe to obtain $a^{(m)}_j$ for each of $a^{(m)} (s')$ using Eq. \eqref{eq:e-a} is, following the treatment of Kolobov and Fabre \cite{kolobov2000quantum}:
\begin{equation}
	\tilde{a}_j^{(m)} = \frac{\tilde{e}_j^{(m)}}{\sqrt{\lambda_j}}. \label{eq:tilde-ae}
\end{equation}
The above equation is based on the fact that the expected values of Eq. \eqref{eq:e-a} lead to $ \langle e_j^{(m)} \rangle = \sqrt{\lambda_j} \langle a_j^{(m)} \rangle$ 
since $\langle c_j^{(m)} \rangle = 0$ for all $j$ because values of $a^{(m)} (s')$ outside the domain $|s'|\leq 1$ is zero (i.e., the object only exists in the range $|s'| \leq 1$, and the coefficients $c_j^{(m)}$ represent quantum noise and thus cancel out when averaged).

\section{Prolate Spheroidal Functions and Eigenvalues}
\label{sec: Prolate Spheroidal Functions and Eigenvalues}

\begin{figure}
	\centering
	\begin{subfigure}[b]{0.9\linewidth}
		\includegraphics[width=0.9\linewidth]{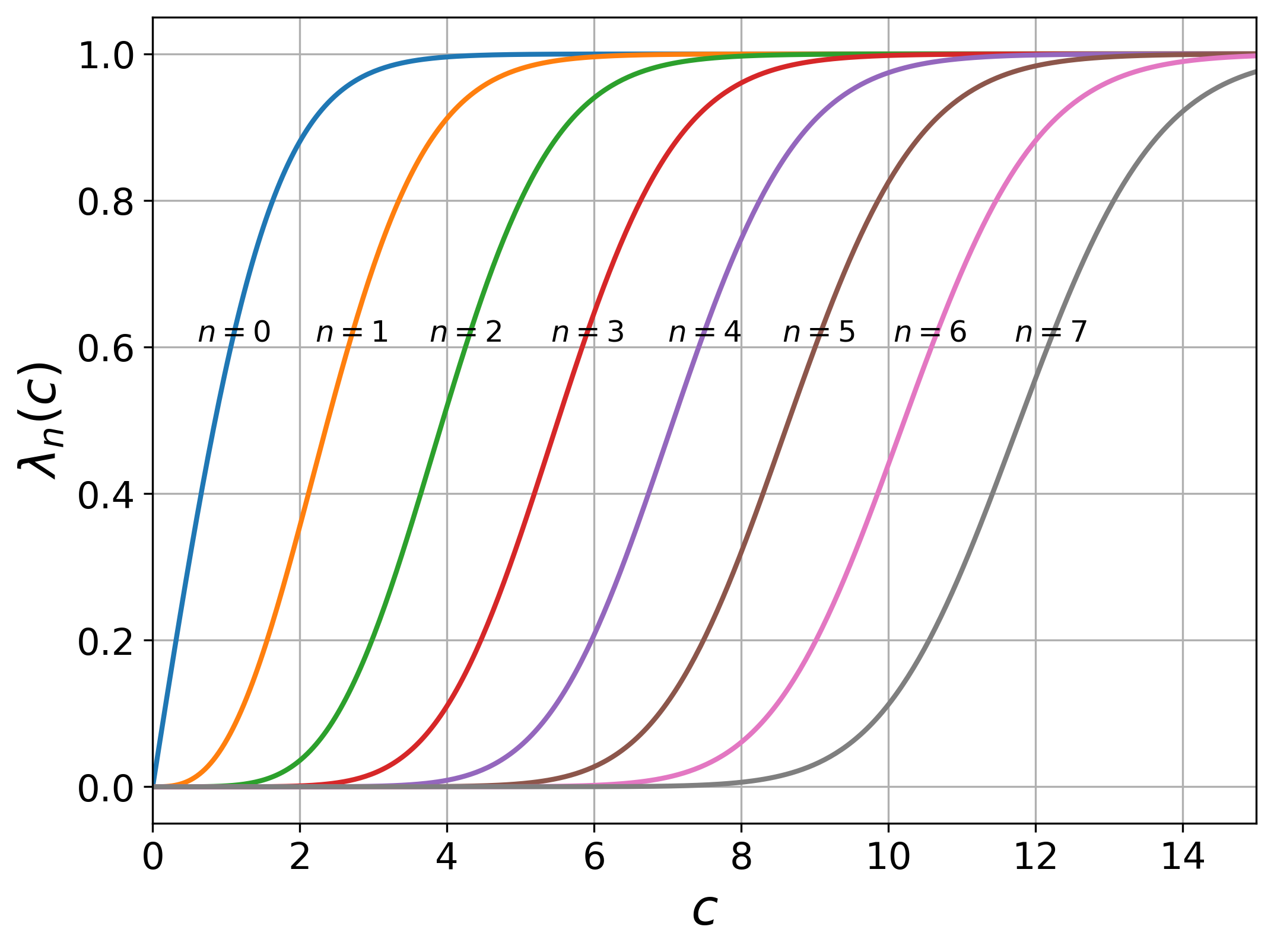}
		\caption{}
	\end{subfigure}
	\begin{subfigure}[b]{0.97\linewidth}
		\includegraphics[width=0.9\linewidth]{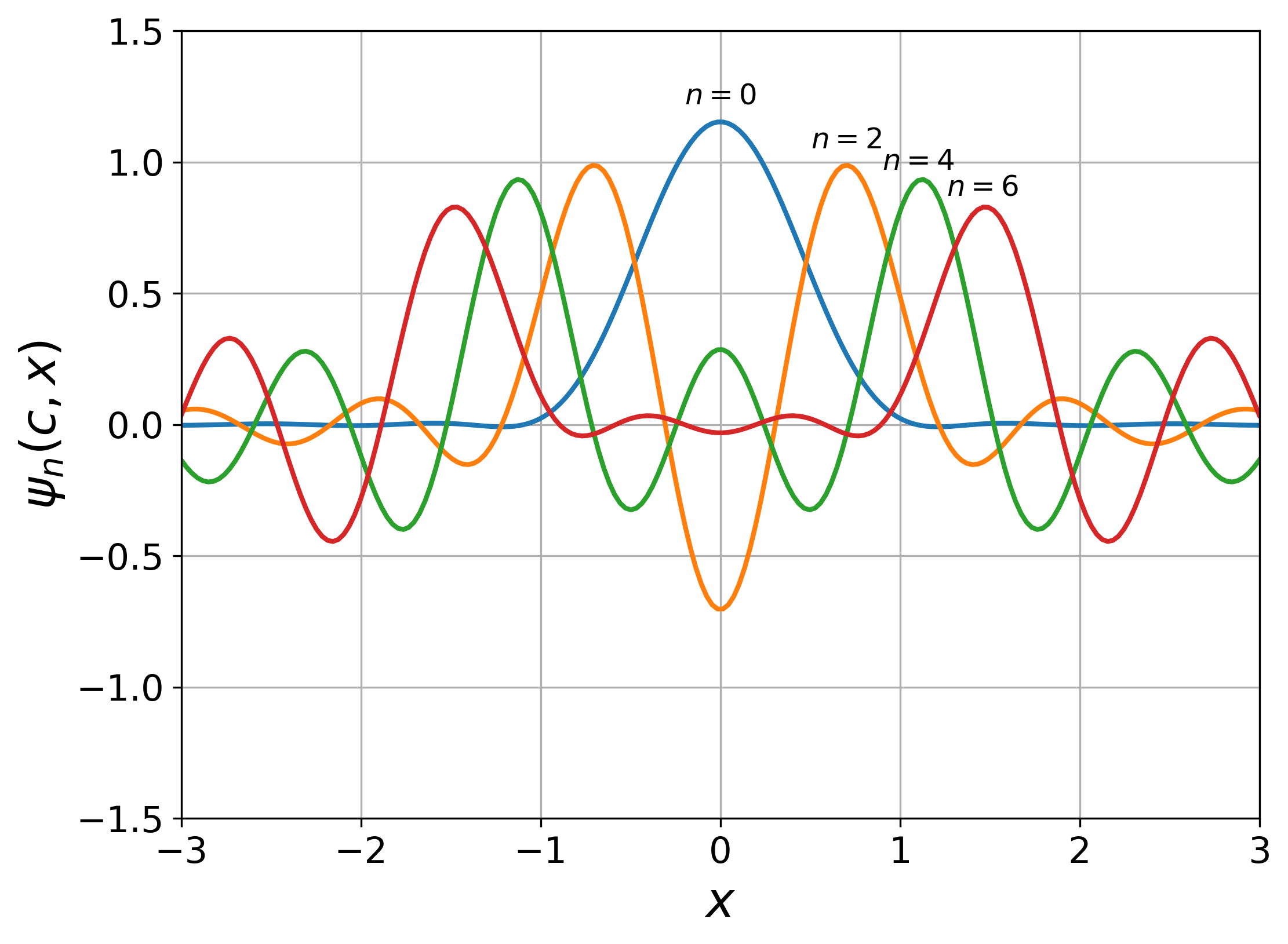}
		\caption{}
	\end{subfigure}
	\begin{subfigure}[b]{0.97\linewidth}
		\includegraphics[width=0.9\linewidth]{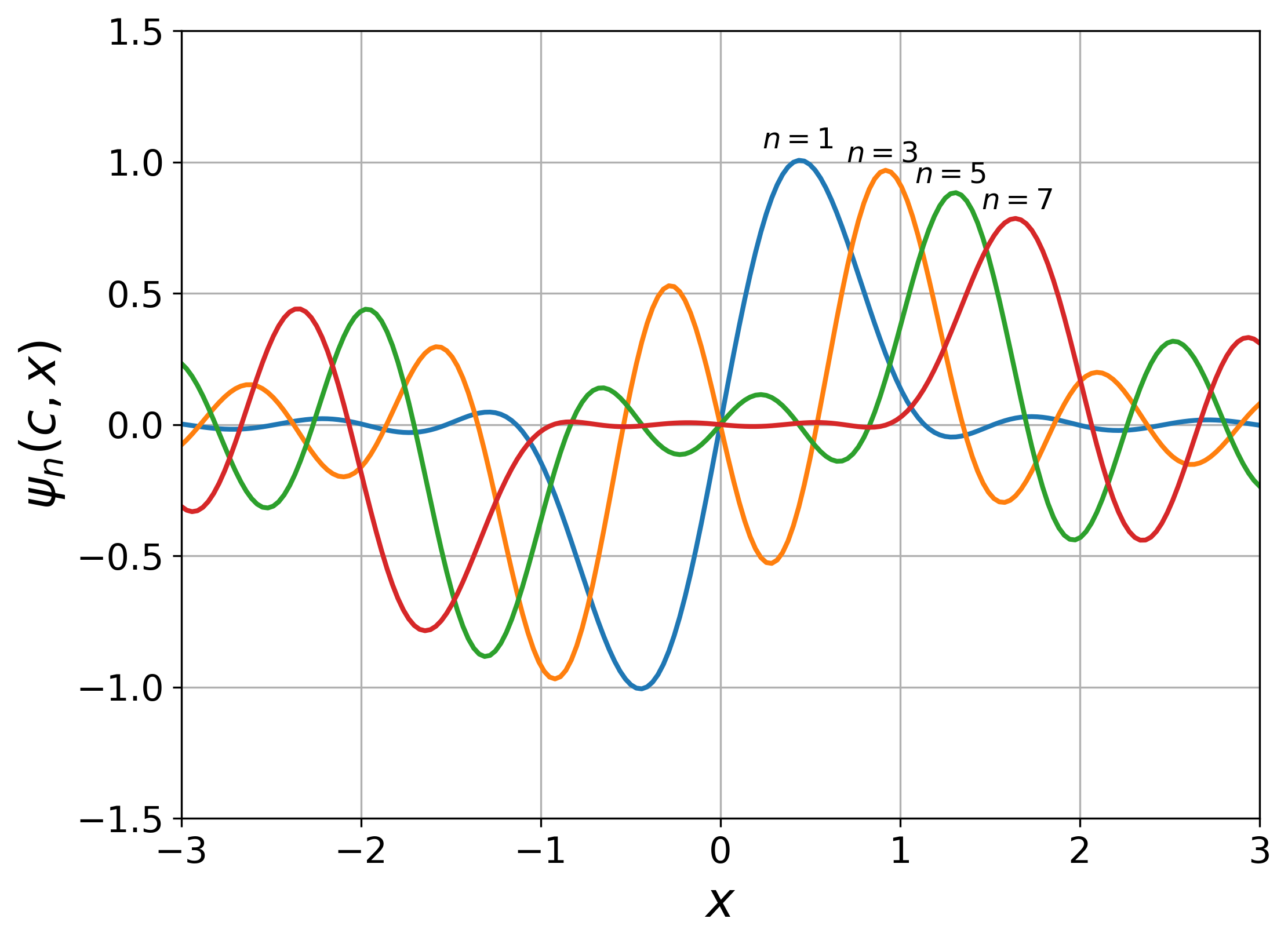}
		\caption{}
	\end{subfigure}
	\caption{Prolate spheroidal functions. (a) Eigenvalues, (b) even orders ($c=6$), and (c) odd orders ($c=6$).}\label{fig:splenian_functions}
\end{figure}

Figure~\ref{fig:splenian_functions} shows the even and odd prolate spheroidal functions, along with the eigenvalues corresponding to the object-to-image transformation for each of the functions.

\section{Derivation of Object-to-Image Transformation}
\label{sec: Derivation of Object-to-Image Transformation}

We now heuristically derive the object-to-image transformation $L$ given the setup in Fig.~\ref{fig:imaging}, starting with an object with the profile $\psi(s')$. In general, for an object size much smaller than the pupil size yet much larger than the light wavelength (i.e., $\lambda \ll Y \ll d$, assuming that the focal length $f$ and pupil size $d$ are on the same order), the size of the image would approximately equal that of the object, with each point on the object focusing onto the corresponding point on the image screen. This is due to the fact that the diffraction-induced spread of the image (on the order of $\lambda f/d$) is neligible compared to the size of the object $Y$. Furthermore, the large size of the object relative to the wavelength implies a narrow beam angle, ensuring that all of the propagating beams are transmitted through both lenses and onto the screen. Therefore, the intensity of the beams arriving on the screen would match that of the beams leaving the object. Consequently, the object-to-image transformation reduces to identity, i.e. $L[\psi](s) = \psi(s)$.

For the case of a very small object (below the wavelength scale), however, the pupil size becomes a limiting factor. Specifically, the object size is now much smaller than the diffraction-induced spread $\lambda f/d$ on the screen, resulting in the image size being significantly larger than the object size (i.e., $L[\psi](s) \neq 0$ for $|s| > 1$). By the same principle, if we consider very closely-spaced points on a given object, then the diffraction patterns corresponding to the different points will overlap and interfere on the image screen. The one-to-one mapping between object and image is thus lost, and a more intricate analysis is required to recreate the object from the image information.

We analyze the image by splitting the object into infinitessimal segments of length $ds'$ and summing the contributions of the beams from each segment onto the point $s$ on the screen. Since a generic segment at point $s'$ on the object axis gives rise to a field of amplitude $\psi(s')ds'$, the overall amplitude of the light focused on the point $s$ is calculated by summing the diffraction patterns using the following integral:
\begin{equation}
L[\psi](s) \propto \int_{-\infty}^{\infty} ds' \psi(s') \frac{\sin{\Big(\frac{d\pi}{\lambda} \sin{(\theta' - \theta)}\Big)}}{\frac{d\pi}{\lambda} \sin{(\theta' - \theta)}},
\end{equation}
where $\theta'$ represents the angle at which the beams originating from the object point $s'$ traveled prior to reaching the pupil, and $\theta$ represents the angle at which the beams reaching the screen point $s$ traveled after diffracting through the pupil. Note from Fig.~\ref{fig:imaging} that any beam traveling through the origin (i.e., through the pupil on the optical axis) at a generic angle $\theta''$ is bent by the second lens such that it becomes parallel to the optical axis afterward. Therefore, given an object size much smaller than the pupil size (i.e., $Y \ll d$), we apply the small-angle approximation to obtain the relationships $\theta = Ys/(2f)$ and $\theta' = Ys'/(2f)$. Substituting these into the above expression for the object-to-image transformation $L$, we find the following:
\begin{equation} \label{eq: object-to-image proportionality}
L[\psi](s) \propto \int_{-\infty}^{\infty} ds' \psi(s') \frac{\sin{(c(s'-s))}}{c(s'-s)},
\end{equation}
where $c = \pi d Y/(2 \lambda f)$ is the previously defined space-bandwidth product.

It is important to note that the fact that the object-to-image transformation is proportional to the above expression rather than equal to it. This is due to the fact that in the small-object limit ($dY \ll \lambda f$), the wide scattering angle of the light from the object and the limited pupil/lens size ensure that only a fraction of the beams actually enter the lens and make it through the aperture and the second lens to the screen. Heuristically, we know that the tangent of the scattering half-angle of the light is on the order of $\lambda/(2\pi Y)$, since the photon momentum along the propagation axis is $h/\lambda$ while the limitation on the transverse object size ensures a transverse momentum uncertainty with a range of about $h/(\pi Y)$. On the other hand, the tangent of the half-angle subtended by the lens is $d/(2f)$. Consequently, the fraction of the light amplitude that passes through the lens is on the order of $\sqrt{\pi d Y/(\lambda f)}$ for a one-dimensional setup. Extrapolating to the case of a two-dimensional setup with circular symmetry, the amplitude attenuation factor thus becomes about $\pi d Y/(\lambda f)$. This is on the same order as the result found by Kolobov and Fabre \cite{kolobov2000quantum}, i.e., $d Y/(2 \lambda f) = c/\pi$. Multiplying the expression in Eq.~\eqref{eq: object-to-image proportionality} by this factor yields the following equality for the object-to-image transformation:
\begin{equation}
L[\psi](s) = \int_{-\infty}^{\infty} ds' \psi(s') \frac{\sin{(c(s'-s))}}{\pi(s'-s)},
\end{equation}
where the term $c$ in the denominator of Eq.~\eqref{eq: object-to-image proportionality} has been replaced with $\pi$.

\section{Calculating the Image's Superposition Coefficients}
\label{sec: Calculating the Image's Superposition Coefficients}

The superposing weight $e_j^{(m)}$ is obtained through 
\begin{equation}
	e_j^{(m)} = \frac{\int_{-\infty}^\infty ds\, e^{(m)}(s) \psi_j(s)}{\int_{-\infty}^\infty ds\, \psi_j^2 (s)},
\end{equation}
because $\psi_j (s)$ are real functions. 

Now, we introduce a linear transformation $L$ such that
\begin{equation}
	L[f](s) = \int_{-\infty}^\infty ds' \frac{\sin[c (s - s')]}{\pi (s - s')} f(s'), \quad -\infty < s < \infty.
\end{equation}
Then, it easily follows that
\begin{equation}
	L[\varphi_j] (s) = \sqrt{\lambda_j} \psi_j (s), \quad L [\chi_j] (s') = \sqrt{1-\lambda_j} \psi_j (s').
\end{equation}
From these, we obtain the following:
\begin{align}
	L[a^{(m)}] (s) &= \sum_{j=0}^\infty \left( a^{(m)}_j L[\varphi_j] (s) + c^{(m)}_j L[\chi_j] (s) \right) \nonumber \\
	&= \sum_{j=0}^\infty \left( \sqrt{\lambda_j} a^{(m)}_j + \sqrt{1 - \lambda_j} c^{(m)}_j \right) \psi_j (s).
\end{align}
From Eq.~\eqref{eq:def-e}, Eq. \eqref{eq:e-a-transform}, and the above expressions, we obtain the following relation:
\begin{equation}
	e^{(m)}_j = \sqrt{\lambda_j} a^{(m)}_j + \sqrt{1 - \lambda_j} c^{(m)}_j, \label{eq:e-a appendix}
\end{equation}
and we can easily show that $[e^{(m)}_j, e^{(m)\dagger}_k] = \delta_{jk}$, satisfying the quantization condition of the bosonic annihilation operators $e^{(m)}_j$.

\section{Derivation of Quantum Noise}
\label{sec: Derivation of Quantum Noise}

Here, we analyze the noise in the reconstructed object field. We define the noise as
\begin{equation}
	\sigma = \sqrt{\int_{-\infty}^{\infty} ds' ~ \langle (\Delta \tilde{\phi} (s'))^2 \rangle},
\end{equation}
which is the root-mean-squared sum of all noises in the reconstructued spatially-dependent phase. It is straightforward to derive
\begin{widetext}
\begin{align}
	e^{(m)}_{j2} & = \mathrm{Im} \left[ \sqrt{\lambda_j} a_j^{(m)} + \sqrt{1 - \lambda_j} c_j^{(m)} \right] \nonumber \\
	&= \mathrm{Im} \left[ \sqrt{\lambda_j} \int_{-1}^1 ds' b^{(m)} e^{i \phi (s') } \varphi_j (s') + \sqrt{1 - \lambda_j} \int_{|s'|>1} ds' b^{(m)} \chi_j (s') \right] \nonumber \\
	&\approx\sqrt{\lambda_j} \left( b_2^{(m)} \int_{-1}^1 ds' \varphi_j (s') + b_1^{(m)} \int_{-1}^1 ds' \phi (s') \varphi_j (s') \right) +  \sqrt{1 - \lambda_j} \int_{|s'|>1} ds' b_2^{(m)} \chi_j (s') \nonumber \nonumber \\
	&= b_2^{(m)} \int_{-\infty}^\infty ds' \psi_j(s) + b_1^{(m)} \int_{-1}^1 ds' \phi (s') \psi_j (s') \label{eq:ej2 appendix}
\end{align}
\end{widetext}
Using the MEMSL's crucial identity in Eq. \eqref{eq:MEMSL-eq}, we obtain
\begin{align}
	&\sum_{m=1}^M e^{(m)}_{j2} = \nonumber \\
	&  b_{e2} \sqrt{M} \int_{-\infty}^\infty ds' \psi_j (s') + b_{e1} \sqrt{M} \int_{-1}^1 ds' \phi  (s') \psi_j (s'). \label{eq:eb appendix}
\end{align}

Now, we apply the Parseval's theorem to calculate the variance of the noise:
\begin{align}
	&\int_{-\infty}^{\infty} ds' \langle (\Delta \tilde{\phi} (s'))^2 \rangle \nonumber \\
    &= \frac{1}{2 \alpha_e^2 M} \sum_{j=0}^Q \frac{\langle \Delta (\sum_{m=1}^M e^{(m)}_{j2})^2 \rangle}{\lambda_j^2} \int_{-\infty}^{\infty} ds' \psi_j^2(s') \nonumber \\
    &= \frac{1}{2 \alpha_e^2 M} \sum_{j=0}^Q \frac{\langle \Delta (\sum_{m=1}^M e^{(m)}_{j2})^2 \rangle}{\lambda_j^2} \nonumber \\
	&\approx \frac{1}{2 \alpha_e^2 M} \sum_{j=0}^Q \frac{M A_j^2 \langle (\Delta b_{e2})^2 \rangle}{\lambda_j^2} \nonumber \\
	&= \frac{e^{-2 r_e}}{8 \alpha_e^2} \sum_{j=0}^Q \frac{A_j^2}{\lambda_j^2}, \label{eq:der1 appendix}
\end{align}
where we defined 
\begin{equation}
	A_j = \left|\int_{-\infty}^\infty ds' \psi_j (s') \right|. \label{eq:Aj appendix}
\end{equation}
Here, the fourth line comes from the conventional approximation to ignore the contribution of the second term in Eq. \eqref{eq:eb appendix} for the variance calculation since $| \phi(s') | \ll 1$ (for example, see Guo \textit{et al.} \cite{guo2020distributed}). More precisely, the approximation becomes well justified with a conservative sufficient condition 
\begin{equation}
	\left|\int_{-1}^1 ds' ~ \phi (s') \right| < \frac{e^{-r_e}}{\left|\int_{-1}^1 ds' \psi_0 (s')\right|}.
\end{equation}

\nocite{*}

\bibliography{ref.bib}

\end{document}